\documentclass[aps,pra,amsmath,amssymb,showpacs,preprint]{revtex4-1}
\usepackage[T1]{fontenc}
\usepackage{amssymb,amsmath}
\usepackage{amsfonts}
\usepackage{graphicx}
\usepackage{amsthm}
\usepackage{color}
\usepackage{mathbbol}
\usepackage{epstopdf}
\usepackage{float}

\usepackage{tikz}
\usepackage{verbatim}

\def\ket#1{|#1\rangle}

\def\bra#1{\langle#1|}
\def\ketbra#1{|#1\rangle\langle#1|}

\def\tr{\mathrm{tr}}

\newcommand{\ml}[1]{\mathcal{#1}}
\newcommand{\Ec}{\ensuremath{\mathcal{E}}}
\newcommand{\idc}{\ensuremath{ \mathrm{Id} }}
\newcommand{\id}{\ensuremath{ \mathcal{I} }}
\renewcommand{\tr}[1]{\ensuremath{  \operatorname{tr}\! \left[ #1 \right]  }}
\newcommand{\trpart}[2]{\ensuremath{ \, \mathrm{Tr}_{#1} \left[ #2 \right] \, }}
\newcommand{\db}[3][]{\mathop{}\! {d_{\mathrm{B}}}^{#1}\left(#2
    \,,#3\right)}
\newcommand{\dens}[2]{\ensuremath{ \vert \, #1 \, \rangle \langle \, #2 \, \vert}} 
\newcommand{\ddens}[1]{\dens{#1}{#1}} 
\newcommand{\est}[1]{\hat{#1}_{\mathrm{est}}  }
\newcommand{\ens}[1]{\ensuremath{ {\left\{ #1 \right\}}} }
\newcommand{\pscal}[2]{\ensuremath{ \langle \, #1 \, \vert  \, #2 \, \rangle}} 
\def\ket#1{|#1\rangle}

\def\bra#1{\langle#1|}
\def\ketbra#1{|#1\rangle\langle#1|}
\newcommand{\vp}{\ensuremath{\varphi}}
\newcommand{\ii}{\ensuremath{ \mathrm{i\,} }}
\newcommand{\e}[1]{\ensuremath{ \mathrm{e}^{#1}}}
\newcommand{\relphantom}[1]{\mathrel{\phantom{#1}}}

\begin{document}
\title{Quantum channel-estimation with particle loss: GHZ versus W states}
\author{Julien Mathieu Elias Fra\"isse$^{1}$  and Daniel Braun$^{1}$ }
\affiliation{$^1$ Eberhard-Karls-Universit\"at T\"ubingen, Institut
  f\"ur Theoretische Physik, 72076 T\"ubingen, Germany}

\begin{abstract} We consider quantum channel-estimation
  for depolarizing channels and phase-flip channels extended
  by ancilla qubits and fed with a GHZ or W state.  After application
  of the channel one or several qubits can be lost, and we calculate
  the impact of the loss on the quantum Fisher information that
  determines the smallest uncertainty 
with which the parameters of
  these channels can be estimated.  
\end{abstract}

\maketitle
%\begin{twocolumn}

\section{ \label{intro}Introduction}
With the rise of quantum information processing the need to precisely
characterize quantum devices has become an important practical
issue. Contrary to classical bits, qubits have a continuous range of
possible states, and their phase coherence is crucial for 
quantum
algorithms. Under the influence of the environment, qubits suffer
decoherence processes which can manifest themselves not only in a flip
of the bit, but also in a loss of phase coherence.  If the
corresponding error rates are small enough, quantum error-correction
can be applied, which, when concatenated, allows 
 one to ultimately perform
meaningful quantum computations. For the development of the hardware
and optimization of the quantum error correction it is, however,
crucial to precisely know the error mechanisms and rates.  This
amounts to performing a ``quantum channel-estimation''.  For a single
qubit, the set of possible quantum channels and their possible
parametrizations are well known by now
\cite{Bengtsson06,Fujiwara99,braun_universal_2014}, 
and quantum channel-estimation amounts hence to estimating these
parameters.   

Quantum channel-estimation is therefore very similar to quantum
parameter estimation (q-pet), where one
tries to estimate the parameters that determine a quantum
state. Q-pet is very well developed, starting with work by Helstrom
and Holevo in the 1970s and  
Braunstein and Caves in the 1990's
\cite{Helstrom1969,Holevo1982,Braunstein94,Braunstein95,braunstein_generalized_1996,Holevo01}. This 
led to the important insight that the 
smallest possible uncertainty of 
the estimation of a parameter is fully determined by the
parameter-dependence of the quantum state.  The quantum
Cram\'er-Rao bound (QCRB) quantifies this smallest uncertainty and generalizes 
corresponding results from classical statistical analysis from the
 1940s \cite{Rao45,Cramer46}. The theoretical framework has been used to
analyze the ultimate possible sensitivity of gravitational wave
observatories \cite{AasiNP2013Short}, Mach-Zehnder and atomic interferometers
\cite{GrossN2010,HollandPRL1993,NapolitanoNJP2010,TaclaPRA2010},
measurements of time \cite{braunstein_generalized_1996}, mass
\cite{Braun11.2,braun_ultimate_2012},  
temperature and chemical potential
\cite{Stace2010,marzolino_precision_2013,Mehboudi2015,Correa15},
parameters of space-time \cite{Ahmadi14,braun_how_2015}, and many more. 

In  addition to the optimization over all possible measurements and
data analysis schemes that is inherent in the QCRB, quantum
channel-estimation allows one to also optimize over the input
state. It has 
long been known that using highly entangled states can enhance the
precision of certain measurements beyond what is possible
classically \cite{Giovannetti2006}, even though measurements exist where
such enhancements do not need entanglement, or the naturally
occuring ``entanglement'' due to the symmetrization of states of
identical particles is enough for improved performance
\cite{fraisse_coherent_2015,luis_nonlinear_2004,BenattiPRA2013}. 

It is also known that the quantum advantage can break down very
rapidly with the smallest amount of decoherence.  For example,
Markovian decoherence, no matter how small, always leads back to the
so-called standard quantum limit of the uncertainty of atomic clocks,
when these are operated with a highly entangled GHZ state
\cite{huelga_improvement_1997}.  From interferometry it is well known
that GHZ 
states are also maximally fragile against loss of qubits: Even the loss
of a single one turns the state into a uniform mixture of the original
pure components in the Hilbert space of the remaining particles, and hence
erases any useful phase information \cite{Huelga97}. 
W states are more robust here, as 
are ``entangled coherent states''  \cite{zhang_quantum_2013}.

One might wonder then, how useful
highly 
entangled states are for measuring decoherence processes themselves. 
Interestingly, it was found by Fujiwara and Imai that the best estimation of a
Pauli quantum channel of a single qudit can be achieved by sourcing a
maximally entangled state into the channel extended by just one more
qudit
 \cite{fujiwara_quantum_2003}. Frey {\em et al.} studied the depolarizing
 channel and considered the performance of several extension schemes
 \cite{frey_probing_2011}, and Collins and coworkers examined the
depolarizing channel  and the phase-flip channel  fed with mixed states
\cite{collins_mixed-state_2013,collins_depolarizing_2015}.  

In the present work we investigate channel
estimation for the depolarizing channnel and the phase-flip channel,
when these channels are extended by using several ancilla qubits.  We
consider GHZ states and W states as input and investigate the 
effect of loss of one or several of the
qubits, both the original one or the ancillas, on the precision with
which the parameters of the channels can be estimated.

\section{Quantum estimation of channels}
\subsection{Channels} 
Let $\ml{B}_1=\ml{B}(\ml{H}_1)$ be the space of bounded linear operators  acting on a first
Hilbert space $\ml{H}_1$ and $\ml{B}_2=\ml{B}(\ml{H}_2)$ the space of bounded linear operators
acting on a  second Hilbert space $\ml{H}_2$. A quantum channel $\Ec$
is a \emph{completely positive trace 
  preserving } (CPTP) convex-linear map $\Ec:\ml{B}_1\to\ml{B}_2$ 
that maps a density matrix (\emph{i.e.}~a positive linear
operator with trace one)  to another density matrix,
$\rho\in\ml{B}_1\mapsto 
\sigma\in\ml{B}_2$.  The condition of complete positivity means that
the channel should be a positive map (\emph{i.e.}~maps positive operators to
positive ones), but also that the extension $
\Ec \otimes \idc$ of the 
channel  to ancillary Hilbert spaces $\ml{H}$, where it acts by the identity
operator, should be a positive map, {\em i.e. }($ \Ec \otimes \idc
)(A) \geq 0$  for any positive operator $A$ in  $\ml{B}(\ml{H}_1
\otimes \ml{H})$,
the space of bounded operator acting on the bipartite system $\ml{H}_1 \otimes \ml{H}$
\cite{nielsen_quantum_2011}. Trace preservation is defined as 
$\tr{\Ec(\rho)}=\tr{\rho}$, and convex linearity as  $\Ec(\sum_i p_i
\rho_i) =\sum_i p_i \Ec(\rho_i) $ for all $p_i$ with $0\le p_i\le 1$
and $\sum_i p_i=1$.
According to Kraus' theorem, a quantum channel can be represented as 
\begin{equation}
\Ec(\rho)=\sum_i E_i \rho E_i^\dagger\,,
\end{equation}
where the Kraus operators
$\lbrace E_i \rbrace$ satisfy $\sum_i
E_i^\dagger E_i =\id_2$, the identity
operator 
on the target Hilbert space $\ml{H}_2$  \cite{Kraus83}.

In the following  we study the two physically important channels
``depolarizing channel'' and  
``phase-flip channel'' for a single qubit \cite{nielsen_quantum_2011}.
The depolarizing channel describes relaxation,
\begin{equation}
\ml{E}_{\mathrm{dep}}(\rho)=p \frac{\id}{2} +(1-p)\rho \,,
\end{equation} 
{\em i.e.}~the qubit is replaced with
probability $p$ (the ``depolarization strength'') by the totally mixed
state.  
Its Kraus decomposition is given by
\begin{equation}
\mathcal{E}_{\mathrm{dep}}(\rho)=\sum_{i=1}^4 E_i \, \rho \, E_i^\dagger\,,
\end{equation}
with the four Kraus operators:
\begin{equation}
E_1=\sqrt{1-3\frac{p}{4}}\, \id \,,\; E_2=\sqrt{\frac{p}{4}}\, X \,,\;E_3=\sqrt{\frac{p}{4}}\,Y \,,\; E_4=\sqrt{\frac{p}{4}}\,Z\,,
\end{equation}
where $X,Y$ and $Z$ are the three Pauli matrices. 

The phase-flip channel has the  Kraus representation
\begin{equation}
\mathcal{E}_{\mathrm{ph}}(\rho)=\sum_{i=1}^2 F_i \, \rho \, F_i^\dagger\,,
\end{equation}
with the Kraus operators
\begin{equation}
F_1=\sqrt{1-p} \,\id \,,\; F_2=\sqrt{p}\, Z \,,
\end{equation}
{\em i.e.} with probability $p$ the phase of the qubit is flipped. 

We also define extensions of these channels by the identity to $n$ ancilla qubits, on
which the channels act through the identity operation. For the
depolarizing channel we have 
\begin{equation}
\mathcal{E}_{\mathrm{dep}}^{(n)}(\rho)=
\left(\mathcal{E}_{\mathrm{dep}}\otimes\idc \cdots \otimes \idc
\right)(\rho)= \sum_{i=1}^4 \Gamma_i \, \rho \, \Gamma_i^\dagger\,, \label{ext.dep}
\end{equation}
where the Kraus operators $\Gamma_i$ of the extended channel are
defined as $\Gamma_i= E_i \otimes \id^{\otimes n}$.

% \cc{
% \begin{equation}
% \Gamma_1=\sqrt{1-\frac{3}{4}p} \, \id \otimes \id^{\otimes n} \;\,; \Gamma_2=\sqrt{\frac{p}{4}} \,  X\otimes \id^{\otimes n} \,,\;\Gamma_3=\sqrt{\frac{p}{4}} \, Y\otimes \id^{\otimes n} \,,\; \Gamma_4=\sqrt{\frac{p}{4}} \, Z\otimes \id^{\otimes n}\,.
% \end{equation}
% }
Similarly, we extend the phase-flip channel to $n$ ancillary qubits by
\begin{equation}
\mathcal{E}_{\mathrm{ph}}^{(n)}(\rho)=\left(\mathcal{E}_{\mathrm{ph}}\otimes\idc \cdots \otimes \idc \right)(\rho)= \sum_{i=1}^2 \Lambda_i \, \rho \, \Lambda_i^\dagger\;\label{ext.deph}
\end{equation}
with the new Kraus operators $\Lambda_i= F_i \otimes \id^{\otimes n}$.

% \cc{
% \begin{equation}
% \Lambda_1=\sqrt{1-p}\, \id \otimes \id^{\otimes n} \;\,; \Lambda_2=\sqrt{p} \,Z \otimes \id^{\otimes n}\,.
% \end{equation}
% }

\subsection{Quantum parameter estimation}

Quantum parameter estimation theory (q-pet)
\cite{Helstrom1969,Holevo1982,Braunstein94,braunstein_generalized_1996} 
provides a lower bound on the variance of an 
unbiased estimator $\est{\theta}$ of a parameter $\theta$ on which a state $\rho(\theta)$
depends. Its importance arises from the facts that ({\em i.}) it is optimized
already over all possible measurements (POVM measurements,
generalizing projective von Neumann measurements \cite{Peres93}), and
  all possible 
data analysis schemes in the form of unbiased estimators 
({\em i.e.}~estimators  that on the average give back the
true
value of the parameter); and ({\em ii.}) the bound is reachable
at least asymptotically, in the limit of an infinite number of measurements.
This so-called
quantum Cram\'er-Rao bound (QCRB) is given by  
 \begin{equation}\label{qfi_inequality0}
 \mathrm{Var}( \est{\theta} ) \geq \frac{1}{M I(\rho(\theta);\theta)}\,,
\end{equation}
with $M$ the number of independent measurements and $I(\rho(\theta);\theta)$ the
 quantum Fisher information (QFI). 
In \cite{Braunstein94} it was
shown that $I(\rho(\theta);\theta)$ is a geometric measure on how much
$\rho(\theta)$ and $\rho(\theta+d\theta)$ differ, where $d\theta$ is
an infinitesimal increment of $\theta$. The QCRB thus offers the physically
intuitive picture that the parameter $\theta$ can be measured the more
precisely the more strongly the state $\rho(\theta)$ depends on
it (see below for a precise definition).
The geometric measure is given by the Bures-distance, 
\begin{equation} \label{db2}
\db{\rho}{\sigma}^2\equiv 2\left(1-\sqrt{F(\rho,\sigma)}\right)\,,
\end{equation}
where the fidelity $F(\rho,\sigma)$ is defined as
$F(\rho,\sigma)=||\rho^{1/2}\sigma^{1/2}||^2_1$, and $||A||_1\equiv {\rm
  tr}\sqrt{AA^\dagger}$ denotes the trace norm \cite{Miszczak09}. With
this, $I(\rho(\theta);\theta)=4 \db{\rho(\theta)}{\rho(\theta+d\theta)}^2/d\theta^2$
\cite{Braunstein94}. 
The Bures-distance is in general difficult to calculate for mixed
states, but for pure states
$\rho(\theta)=\ket{\psi(\theta)}\bra{\psi(\theta)}$, the QFI reduces to the
overlap of the derivative of the state, 
$\ket{\partial_\theta\psi(\theta)}$, with itself and the original state,
$I(\rho(\theta);\theta)=4(\pscal{\partial_\theta\psi(\theta)}{\partial_\theta\psi(\theta)}-\vert \pscal{\partial_\theta\psi(\theta)}{\psi(\theta)}\vert ^2)$
\cite{Paris09}. 
\\ When the state is not pure we can still give a closed formula by using the spectral representation of the state. For 
\begin{equation}
\rho(\theta) = \sum_{i=1}^d p_i \dens{\psi_i}{\psi_i}\,,
\end{equation}
the QFI can be written as
\begin{equation}
I(\rho(\theta);\theta)=\sum_{\substack{i=1\\p_i\neq 0}}^d \frac{(\partial_\theta p_i)^2}{p_i} +2\sum_{\substack{i,j=1\\p_i+p_j \neq 0}}^d\frac{(p_i-p_j)^2}{p_i+p_j} \vert\pscal{ \psi_j}{\partial_\theta \psi_i}\vert^2\,,
\end{equation}
where the first term is called classical contribution and the second quantum contribution.

The QFI obeys the ``monotonicity property'' under $\theta$-independent channels $\mathcal{E}$
\begin{equation} \label{eq:mon}
  I(\mathcal{E}(\rho(\theta));\theta)\leq I(\rho(\theta);\theta)\,,
\end{equation}
with equality for unitary channels $\mathcal{U}$, defined by $\mathcal{U}(\rho)=U\rho\,
U^\dagger$ \cite{Petz_monotone_1996} with $U$ unitary. 
The QFI has also the property of convexity, meaning that for two
density matrices $\rho(\theta)$ and $\sigma(\theta)$ and  $0\leq
\lambda \leq 1$ we have \cite{fujiwara_quantum_2001} 
\begin{equation}\label{eq:convexity_QFI}
I(\lambda \rho(\theta) +(1-\lambda) \sigma(\theta); \theta) \leq \lambda  I(\rho(\theta); \theta)+(1-\lambda) I( \sigma(\theta); \theta)\,.
\end{equation}
A last useful property of the QFI is the additivity:
\begin{equation}\label{eq:additivity_qfi}
I(\rho(\theta)\otimes \sigma(\theta);\theta )=I(\rho(\theta);\theta )+I( \sigma(\theta);\theta )\,.
\end{equation}
For a state that depends on several parameters $\mathbf{\theta}=(\theta_1,\ldots,\theta_n)$,
the QCRB generalizes to an inequality for the co-variance matrix of
the estimators of the $\theta_i$, with a lower bound 
given by the inverse
of the quantum Fisher information  matrix.  In contrast to the single
parameter case, this inequality can in general not be saturated (see
\cite{szczykulska_multi-parameter_2016} and references therein). 

%%%%%%%%%%%%%%%%%%%%%%%%%%%%%%%%%%%%%%%%%%%%
\subsection{Quantum channel-estimation}
%%%%%%%%%%%%%%%%%%%%%%%%%%%%%%%%%%%%%%%%%%%%

\subsubsection{General considerations}
We consider channels $\mathcal{E}_\theta$ depending on a scalar
parameter $\theta$, and perfectly known initial states $\rho$
independent of $\theta$. After the evolution of $\rho$ through the
channel,  we obtain a state parametrized by $\theta$ with QFI %  of which we can
% compute the QFI to see how much information we can obtain about
% $\theta$: 
\begin{equation}
I(\rho(\theta);\theta)=I(\mathcal{E}_\theta(\rho); \theta)\;
\end{equation} 
that can still be optimized over $\rho$. 
Due to the convexity of the
QFI, its maximal value can be achieved with a pure state.  
The fact that a quantum channel is a completely positive map 
allows one to extend it to a larger Hilbert space by acting with an
arbitrary quantum channel $\mathcal{A}$ on the Hilbert space of the ancilla,
\begin{equation}
\mathcal{E}_\theta^{\mathrm{ext},\mathcal{A}}=\mathcal{E}_\theta \otimes \mathcal{A}\,.
\end{equation}
According to an argument by Fujiwara \cite{fujiwara_quantum_2001}, the
largest QFI with a parameter independent $\mathcal{A}$ can be achieved 
already by choosing for $\mathcal{A}$ the identity channel in the
ancillary Hilbert 
space: Since $\mathcal{E}_\theta \otimes \mathcal{A}$ 
can be decomposed as 
$(\idc \otimes  \mathcal{A})(\mathcal{E}_\theta \otimes
\idc)$, monotonicity of the QFI implies that the best choice for
$(\idc \otimes  \mathcal{A})$ is a unitary channel,
which is the case when $ \mathcal{A}$ is a unitary channel. 
The simplest solution consists in
 taking the identity channel and thus, in the following, when we refer to extensions, we always mean an extension \emph{by the identity
  channel in the Hilbert space of the ancilla}.

% \cc{Channel estimation hence amounts here to quantum
% parameter 
% estimation with the additional freedom of optimizing over
% the input state \cite{fujiwara_quantum_2001}.  
% The estimation  is done by choosing
% an input state  
% (which is supposed to be perfectly known), let it undergo the
% evolution through the channel, and then make some measurement on the final
% state in order to infer the value of the parameter. \\ 
% Both depolarizing and phase-flip channels are parameterized by just
% one parameter, $p$.  }

\subsubsection{Estimation of depolarizing and phase-flip channels}
 
 For both depolarizing and phase-flip channels, the parameter to be
 estimated is $p$. To avoid cumbersome notation, we omit the
 dependence on $p$ in the states, the channels, and the QFI.

For the depolarizing channel acting on one qubit, all states related
by a $p$-independent unitary transformation $U$
 give rise to the same QFI, as
\begin{equation}
\mathcal{E}_{\mathrm{dep}}(U \rho U^\dagger)=U \mathcal{E}_{\mathrm{dep}}(\rho) U^\dagger\,,
\end{equation}
coupled to the fact that the QFI is invariant under
parameter-independent unitary transformations of the state.
For the phase-flip channel, the initial state and in particular the
orientation of its Bloch vector  matters, as in general
for an arbitrary unitary $U$
\begin{equation}
\mathcal{E}_{\mathrm{ph}}(U \rho U^\dagger) \neq U \mathcal{E}_{\mathrm{ph}}(\rho) U^\dagger\,.
\end{equation}

\subsection{Known results}
In \cite{fujiwara_quantum_2003} Fujiwara and Imai
investigated the problem of estimating generalized Pauli 
channels acting on qudits --- \emph{i.e.} systems with a
$d$-dimensional Hilbert space and, in general, $d^2-1$
parameters $\ens{p_i}_{1\leq i\leq d^2-1}$ to estimate. 
The authors were interested in the optimal protocol for estimating
these parameters when one uses the
channel $m$ times. They showed that the optimal protocol
(in terms of 
the QFI matrix) consists in making $m$ independent estimations
of the channel extended to a single ancillary qudit with the same
dimension of Hilbert space 
and to input a pure, maximally entangled state $\ket{\psi^{\mathrm{m.e.}}_d}$
\begin{equation}\label{eq:def_max_ent_states}
\ket{\psi^{\mathrm{m.e.}}_d}=\sum_{i=1}^d \frac{1}{\sqrt{d}}\ket{u_i}\otimes\ket{v_i}
\end{equation}
with $\ens{\ket{u_i}}$ and $\ens{\ket{v_i}}$ two orthonormal bases ($\pscal{u_i}{u_j}=\pscal{v_i}{v_j}=\delta_{ij}$).

 In the specific case of the qubit ($d=2$), the  Pauli channels are the 
channels constructed with Pauli matrices as Kraus operators, 
\begin{equation}
\Ec_{\mathrm{Pauli}}(\rho)=(1-p_1-p_2-p_3)\rho+p_1 X\rho X + p_2 Y\rho
Y+p_3 Z \rho Z\,, 
\end{equation}
{\em i.e.} the estimation of Pauli channels for qubits is in general a 3-parameter
estimation problem.  It reduces to the estimation of the depolarizing
channel by setting $p_1=p_2=p_3=p/4$, while the
phase-flip channel corresponds to the case $p_1=p_2=0$. 
The well-known four Bell states\begin{equation}\label{eq:def_bell_states}
\ket{\phi_\pm}=\frac{\ket{00}\pm \ket{11}}{\sqrt{2}}\,, \ket{\vp_\pm}=\frac{\ket{01}\pm \ket{10}}{\sqrt{2}}\,,
\end{equation} are special cases of maximally entangled states for $d=2$, and
thus achieve the optimal QFI for the estimation of $\lbrace
p_1,p_2,p_3 \rbrace$, the three parameters attached to Pauli
channels for qubits.

  % : 
% \begin{equation}\label{eq_best_est}
% (\Ec \otimes \id)^{\otimes m}({\rho_{\mathrm{max. ent}}}^{\otimes m}) \; , \; \rho_{\mathrm{max. ent}}=\frac{1}{\sqrt{d}}\sum_i^d \ket{e_i}\otimes \ket{t_i}
% \end{equation}
% where $\lbrace \ket{e_i} \rbrace$ and $\lbrace \ket{t_i} \rbrace$ are
% orthonormal basis for the probe and the ancilla.  % (iid. strategy ---
%                                 % standing for \emph{independent
%                                 % identical distributed} ) 

Frey \emph{et al.} analyzed the depolarizing channel
acting on qudits 
 using different extension schemes, including
sequential protocols, where the same probe undergoes $m$ times the
channel before any measurement is done.  \cite{frey_probing_2011}. 
A fair figure of merit for the comparison is then the QFI \emph{per
  channel application}. 
The schemes studied were
({\em i\,}) the non-extended original channel $\Ec_{\mathrm{dep}}$; 
({\em ii\,}) the channel extended by the identity in an ancillary
$q$-dimensional Hilbert space, 
$\Ec_{\mathrm{dep}}\otimes \idc$;
({\em iii\,}) the original channel applied in parallel to two
different qudits, 
$\Ec_{\mathrm{dep}}\otimes \Ec_{\mathrm{dep}}$;
({\em iv\,}) the channel extended by a known depolarizing channel with
depolarizing strength $\eta$, $\Ec_{\mathrm{dep},p}\otimes
\Ec_{\mathrm{dep},\eta}$, where the subscripts $p$ and $\eta$ denote
the respective depolarizing strengths,
and
({\em v\,}) the $m$ times iterated
 use of the channels in schemes ({\em i,
  ii, iii\,}).
Pure input states were considered, with a maximally entangled state   in all
the schemes with more than one qudit, and in addition partially
entangled states in scheme ({\em ii\,}). 

From the work of Fujiwara and Imai \cite{fujiwara_quantum_2003} it is
clear that the best scheme is ({\em ii\,}) 
with a maximally entangled state as input. It also turns out that the
multiple use of the probes is useless in the sense that the QFI per
channel use is always smaller or equal in the sequential schemes
than in the non-sequential
 ones. 
Depending on the dimension $d$ of the Hilbert space of the qudit, 
 and on the depolarization strength $p$, the simple scheme ({\em i\,}) or the double use of
the channel ({\em iii\,}) fair better. 
Partially entangled pure states in scheme ({\em ii\,}) lead to a QFI lying between the one of the optimal scheme and the one of the simple scheme. 
When the additional depolarization $\eta$ in scheme ({\em iv\,})  is too large, it becomes
more efficient to just use the simple channel or the doubled channel.

Collins considered mixed states for the estimation of the phase-flip
 channel \cite{collins_mixed-state_2013} and in
 \cite{collins_depolarizing_2015} Collins and Stephens did the same for
the depolarizing channel.  
They studied sequential protocols, where there is just one qubit
available on which the channel is applied $m$  times, and also parallel
or multi-qubit protocols. For these they investigated the effect of
correlation among more than two qubits on the efficiency. Again the
figure of merit was the QFI per channel application, and the results
were compared to the protocol with just one channel and one qubit
(SQSC protocol). Depending on purity and depolarization strength, both
sequential and correlated 
protocols can outperform the SQSC protocol. Especially for extremely
small purity of qubits, adding more ancillas in the correlated
protocol increases the QFI, and the correlated protocol proves to be better
than the sequential one.

\section{Benchmarks and strategies} 

Fujiwara and Imai's optimal  metrological strategy for the estimation of
  Pauli channels implies that no gain in the QFI is 
  to  be expected by extending the channels to ancillary 
Hilbert spaces with a
  dimension greater than the one of the original space. Nevertheless,
such extensions still have an interest 
  in the case where one faces the loss of
  particles. In this non-ideal situation, adding more ancillas to the
  probe 
  may 
eventually prove useful. We thus 
  study channel estimation with 
W and GHZ states composed of $n+1$ qubits (the original
  probe and $n$ ancillas) as input, and investigate in particular the
  robustness of these schemes under loss of particles.

\subsection{Benchmark}\label{sec.benchmark}
We first calculate two benchmarks for the QFI: The first one, 
$I^{\mathrm{opt}}_{\mathcal{E_\theta}}$, 
corresponds to the 
optimal case identified by Fujiwara and Imai
\cite{fujiwara_quantum_2003}, namely extending the quantum
channel  by the identity to a second qubit and feeding it with a
maximally entangled state:
$I^{\mathrm{opt}}_{\mathcal{E_\theta}}=I((\mathcal{E_\theta}\otimes
\idc) (\ddens{\psi^{\mathrm{m.e.}}});\theta)$.   

The second one, $I^{\mathrm{sep}}_{\mathcal{E_\theta}}$, is given by directly
estimating the parameter of the channel acting on a single qubit and
optimizing over all pure input states: $I^{\mathrm{sep}}_{\mathcal{E_\theta}} = \max_{\ddens{\psi}} I(\mathcal{E_\theta}(\ddens{\psi});\theta)$.  This latter
scheme is, in terms of QFI, equivalent to the case where 
one uses an extended channel (of the form \eqref{ext.dep} or
\eqref{ext.deph}, or in fact an arbitrary $\theta$-independent extension acting separately
on the original system and the ancillas) but inputs a 
separable state. Indeed, due to the additivity of the QFI we have 
\begin{equation}
I(\left(\mathcal{E_\theta}\otimes \mathcal{A}\right)(\rho \otimes \sigma);\theta) =I(\mathcal{E_\theta}(\rho )\otimes \mathcal{A}(\sigma);\theta)= I(\mathcal{E_\theta}(\rho);\theta)+I(\mathcal{A}(\sigma);\theta)=I(\mathcal{E_\theta}(\rho);\theta)\,,
\end{equation}
since the state $\mathcal{A}(\sigma)$ is $\theta$-independent. Thus we refer to this case as 
``separable strategy''.

\subsubsection{Depolarizing channel}
In the non extended case, the QFI for the depolarizing channel depends
only on the purity of the 
input state. When starting with a pure state of a single qubit we
obtain for the QFI  
\begin{equation} \label{eq:qfi_dep_sep}
I_{\mathrm{dep}}^{\mathrm{sep}}=\frac{1}{p(2-p)}\,.
\end{equation}
The optimal strategy leads to
\begin{equation}\label{eq:qfi_dep_bell}
I_{\mathrm{dep}}^{\mathrm{opt}}=\frac{3}{p(4-3p)}\,.
\end{equation}
In more detail, the depolarizing
channel transforms a Bell state  as 
\begin{equation}
\mathcal{E}_{\mathrm{dep}}^{(1)}(\phi_+)=(1-3\frac{p}{4})\phi_+ + \frac{p}{4}(\phi_-+\vp_++\vp_-)\,,
\end{equation}
where $\phi_\pm=\ddens{\phi_\pm}\,,
\vp_\pm=\ddens{\vp_\pm}$. \emph{I.e.}~the channel creates a mixture
between $\phi_+$ and a 
state orthogonal to it, $\phi_-+\vp_++\vp_-$
\cite{preskill_lecture_note}. This makes the scheme 
more sensitive to the value of the parameter than for the separable strategy. \\
With the two benchmarks (\ref{eq:qfi_dep_sep},\ref{eq:qfi_dep_bell}) we
can check whether extending the 
  channel 
  still leads to an improvement compared to the separable strategy
  when qubits can be lost 
by  comparing the QFI to $I_{\mathrm{dep}}^{\mathrm{sep}}$,
 but also how far the QFI is stable against loosing qubits
  %how far loosing qubits reduces the QFI
   compared to the optimal
  strategy, a property that we call ``robustness''.

\subsubsection{Phase-flip channel}
The case of the phase-flip channel is slightly different. Due to the
anisotropy of the channel,  the QFI of the non-extended strategy
depends on the polar angle $\theta$ of the Bloch vector.  The QFI is
optimized by states
$\ket{\psi_{xy}}=(\ket{0}+\e{- \ii  
  \varphi} \ket{1})/\sqrt{2}$ (i.e.~$\theta=\pi/2$), and has the value 
\begin{equation}\label{eq:qfi_ph_sep}
I_{\mathrm{ph}}^{\mathrm{sep},xy}=\frac{1}{p(1-p)}\,.
\end{equation}
The optimal strategy 
 leads to  
\begin{equation}\label{eq:qfi_ph_bell}
I_{\mathrm{ph}}^{\mathrm{opt}}=\frac{1}{p(1-p)}\,,
\end{equation}
which is equal to $I_{\mathrm{ph}}^{\mathrm{sep},xy}\equiv
I_{\mathrm{ph}}^{\mathrm{sep}}$, showing that the state
$\ket{\psi_{xy}}$ is optimal for the separable strategy (since
$I_{\mathrm{ph}}^{\mathrm{opt}}$ 
is an upper bound for the QFI of the separable strategy, and this upper bound is reached with the states $\ket{\psi_{xy}}$).
 For ideal phase-flip channels the extension is thus useless, in the sense that we can achieve the same sensitivity with separable states or entangled ones \footnote{This is a well known fact. Indeed, in \cite{fujiwara_quantum_2003}(part 4.) the authors emphasized that when estimating the parameter of a Pauli channel lying on the boundaries of the tetrahedron of the simplex representing the different Pauli channels, non-maximally entangled states may be as efficient as maximally entangled ones. They also noticed that for the phase damping channel and for the bit flip channel a separable state is already optimal.}.
Here both benchmarks \eqref{eq:qfi_ph_sep} and \eqref{eq:qfi_ph_bell}
are equal. Hence there is no metrological interest in adding any
ancillas. Nevertheless, from a mathematical perspective it is still
interesting to see the effect of adding ancillas and loosing a
fraction of them. 

\subsection{Used input states}

For feeding our extended channels we consider two kinds of entangled states, GHZ states and W states.
The GHZ (Greenberger-Horne-Zeilinger) state for $n+1$ qubits is 
defined as 
\begin{equation}
\ket{\psi^{\text{GHZ-}n}}= \frac{1}{\sqrt{2}}( \ket{0,\mathbb{0}_n}+\ket{1,\mathbb{1}_n})\,, 
\end{equation}
with $\ket{0,\mathbb{0}_n}=\ket{0}_1 \otimes \ket{0}_2 \otimes  \cdots
\otimes \ket{0}_{n+1}$,  $\ket{0,\mathbb{1}_n}=\ket{0}_1 \otimes
\ket{1}_2 \otimes \cdots \otimes \ket{1}_{n+1}$,
$\ket{1,\mathbb{0}_n}=\ket{1}_1 \otimes \ket{0}_2 \otimes  \cdots
\otimes \ket{0}_{n+1}$ and $\ket{1,\mathbb{1}_n}=\ket{1}_1 \otimes
\ket{1}_2 \otimes \cdots \otimes \ket{1}_{n+1}$. Here and in the
following, the first Hilbert space is the one of the probe and all the
others are for ancillas. Here and in the 
following we take $n\geq 1$. 
When $n=1$, the GHZ state $\ket{\psi^{\text{GHZ-}1}}$  is equal to the
Bell state $\ket{\phi_+}$. 
GHZ states are very prone to decoherence,
 in the sense that if even 
a single qubit is lost (traced out), we end up with a mixed
non-entangled state (see eq.\eqref{ghzmix} below).
 We define  the
density matrix
$\rho^{\text{GHZ-}n}=\ketbra{\psi^{\text{GHZ-}n}}$. \\

The W  state  for $n+1$ qubits, W-$n$ for short, is defined as
\begin{equation}
\ket{\psi^{\text{W-}n}}= \frac{1}{\sqrt{n+1}} \sum_{i=1}^{n+1} \ket{1_i}, 
\end{equation}
with  $\ket{1_i}=\ket{0}_1 \otimes \cdots  \otimes\ket{0}_{i-1}
\otimes\ket{1}_i  \otimes \ket{0}_{i+1} \otimes \cdots  \otimes
\ket{0}_{n+1}$, $\forall i \in \{1,\cdots,n+1\}$, \emph{i.e.}~it corresponds
to a single excitation distributed evenly over all qubits. The
case $n=1$ gives also a Bell state: $\ket{\psi^{\text{W-}1}}=\ket{\vp_+}$.  
We also define
$\rho^{\text{W-}n}=\ketbra{\psi^{\text{W-}n}}\,.$

\section{Estimation of the ideal quantum channels}
We start with the situation where no qubits are lost, and determine
the QFI for 
both GHZ and W states for the two channels that we
are interested in. 

\subsection{Depolarizing channel}

\subsubsection{GHZ states }

For the depolarizing channel acting on the GHZ state, we define
\begin{align}
\rho_{\mathrm{dep}}^{\text{GHZ-}n}& \equiv
\mathcal{E}_{\mathrm{dep}}^{(n)} (\rho^{\text{GHZ-}n}) \\
&=\frac{2-p}{4} \left( \dens{0,\mathbb{0}_n}{0,\mathbb{0}_n} +\dens{1,\mathbb{1}_n}{1,\mathbb{1
}_n}\right)+\frac{1-p}{2} \left( \dens{1,\mathbb{1}_n}{0,\mathbb{0}_n}\right. \nonumber\\ 
&\relphantom{=}{}+ \left.\dens{0,\mathbb{0}_n}{1,\mathbb{1}_n}  \right) +\frac{p}{4} \left(  \dens{1,\mathbb{0}_n}{1,\mathbb{0}_n}+\dens{0,\mathbb{1}_n}{0,\mathbb{1}_n} \right)\,.
\end{align}
The density matrix has rank four for $n\geq 1$ (while for $n=0$ it has
rank 2), 
 but eigenvalues and  eigenvectors
are still found easily,
\begin{equation}
\left\lbrace
\begin{aligned}
\sigma^{\mathrm{dep}}_1&=\frac{p}{4}\quad, \quad \sigma^{\mathrm{dep}}_2=\frac{p}{4} \quad,  \sigma^{\mathrm{dep}}_3=1-\frac{3p}{4} \quad, \quad \sigma^{\mathrm{dep}}_4=\frac{p}{4}  \\
\ket{s^{\mathrm{dep}}_1} &= \ket{0,\mathbb{1}_n} \quad,\quad \ket{s^{\mathrm{dep}}_2} = \ket{1,\mathbb{0}_n}\quad,\quad \ket{s^{\mathrm{dep}}_3} = \frac{1}{\sqrt{2}}\left( \ket{0,\mathbb{0}_n} +\ket{1,\mathbb{1}_n} \right)\\ 
\ket{s^{\mathrm{dep}}_4} &= \frac{1}{\sqrt{2}}\left( \ket{0,\mathbb{0}_n} -\ket{1,\mathbb{1}_n} \right)\,.
\end{aligned}\right.
\end{equation}

The eigenvectors are independent of $p$, and the QFI reduces to
its classical part, 
\begin{equation}
I_{\mathrm{dep}}^{\text{GHZ-}n}=\frac{3}{p(4-3p)}= I_{\mathrm{dep}}^{\mathrm{opt}}\,.
\end{equation}
The QFI is independent of the number of
ancillas for $n\ge 1$ and equals the QFI corresponding to the optimal case.

\subsubsection{W states}

For the depolarizing channel and the W states we have
\begin{align}
\rho_{\mathrm{dep}}^{\text{W-}n} & \equiv \mathcal{E}_{\mathrm{dep}}^{(n)}(\rho^{\text{W-}n})\\
&=\frac{p}{2(n+1)} \left(\dens{0,\mathbb{0}_n}{0,\mathbb{0}_n}+ \sum_{i=2}^{n+1}\dens{1,1_i}{1,1_i}+  \sum_{\substack{ i,j=2 \\ i \neq j }}^{n+1} \dens{1,1_i}{1,1_j} \right) \nonumber \\
 & \relphantom{=} {}+\frac{2-p}{2(n+1)} \left(\sum_{i=1}^{n+1}  \dens{1_i}{1_i}+ \sum_{\substack{ i,j=2 \\ i \neq j }}^{n+1}  \dens{1_i}{1_j} \right)+\frac{1-p}{n+1} \sum_{i=2}^{n+1}(\dens{1_i}{1_1}+\dens{1_1}{1_i})\,,   
\end{align}
with $\ket{1,1_i}=\ket{1}_1 \otimes\ket{0}_2\otimes \cdots
\otimes\ket{0}_{i-1} \otimes\ket{1}_i  \otimes \ket{0}_{i+1} \otimes
\cdots  \otimes \ket{0}_{n+1}$, $\forall i \in \{2,\cdots,n+1\}$. 
The matrix representation in the computational basis has a block
structure whose blocks are studied in the appendix, with three non
zero blocks:  
\begin{itemize}
\item a first trivial 1$\times 1$ block composed  by the eigenvalue
  $\frac{p}{2(n+1)}$. 
\item a second block $G^{(n)}(a)$ with $a=\frac{p}{2(n+1)}$.
\item a third block $K^{(n+1)}(a,b,a)$ with $a=\frac{2-p}{2(n+1)}$ and $b=\frac{1-p}{n+1}$.
\end{itemize} 
This leads to the QFI
\begin{equation}\label{eq:qfi_dep_w_n}
I_{\mathrm{dep}}^{\text{W-}n}=\frac{1}{p (2-p)} \frac{\left(3p -4(1+n(n+4))/(1+n)^2\right)}{(3p-4)}\,.
\end{equation}
Even if this
  analysis is restricted to $n\geq 1$, the
  eq.\eqref{eq:qfi_dep_w_n} for $n=0$ gives the correct QFI.
We notice that $I_{\mathrm{dep}}^{\text{W-}n}$ decreases as function
of $n$, {\em i.e.}~adding ancillas reduces the efficiency of the scheme
(see right plot in Fig.\ref{fig:QFI_dep_loss_1}). %Same remark.
 When we go to an infinite number of ancillas, 
\begin{equation}\label{eq:qfi_dep_ideal_W}
I_{\mathrm{dep}}^{\text{W-}n}\; \underset{n\rightarrow \infty}{\longrightarrow}  \;\frac{1}{p(2-p)}=I_{\mathrm{dep}}^{\mathrm{sep}}\,,
\end{equation}
{\em i.e.}~we come back to the case without ancilla.

\begin{figure}
\includegraphics[scale=0.25]{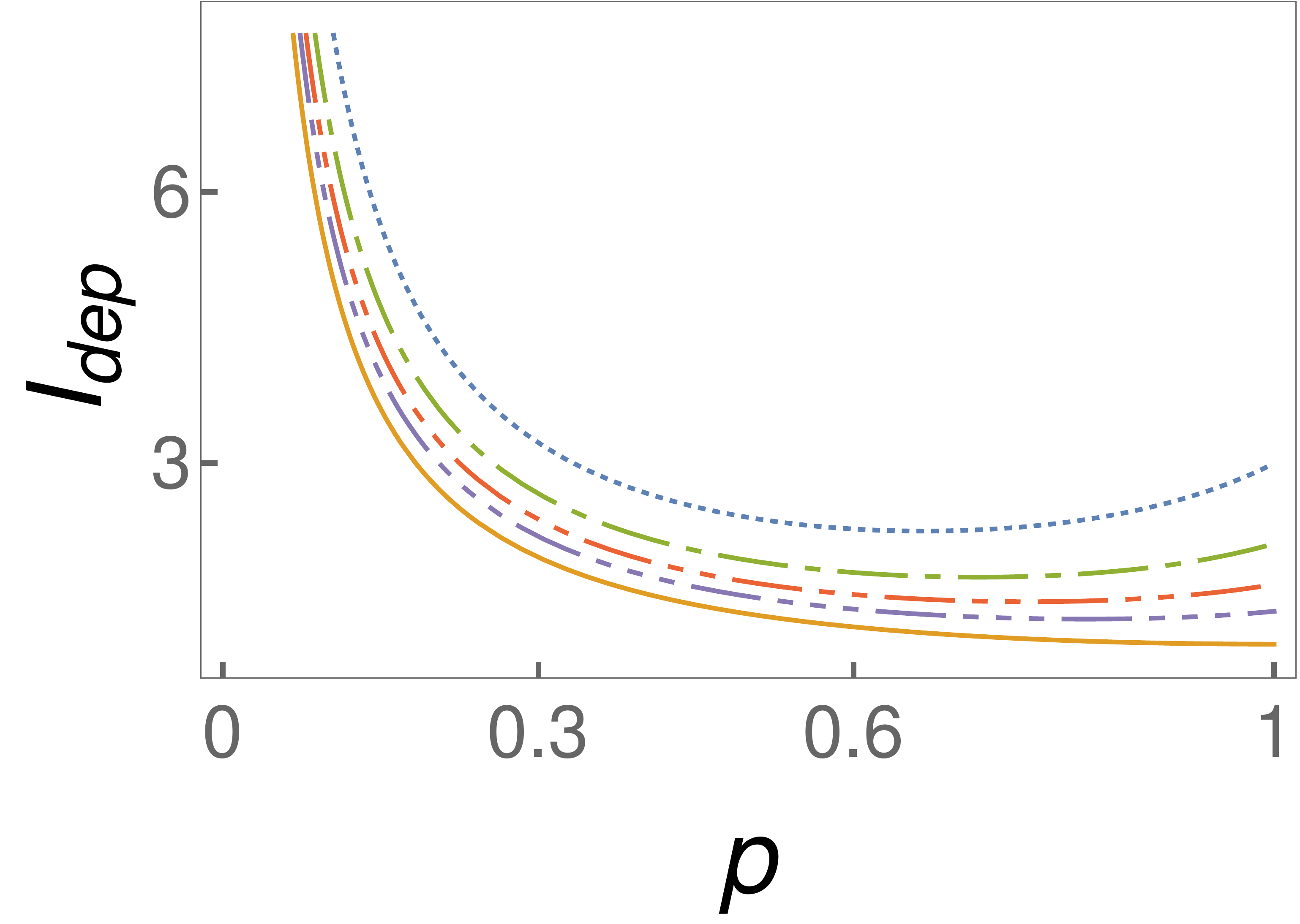}
\includegraphics[scale=0.25]{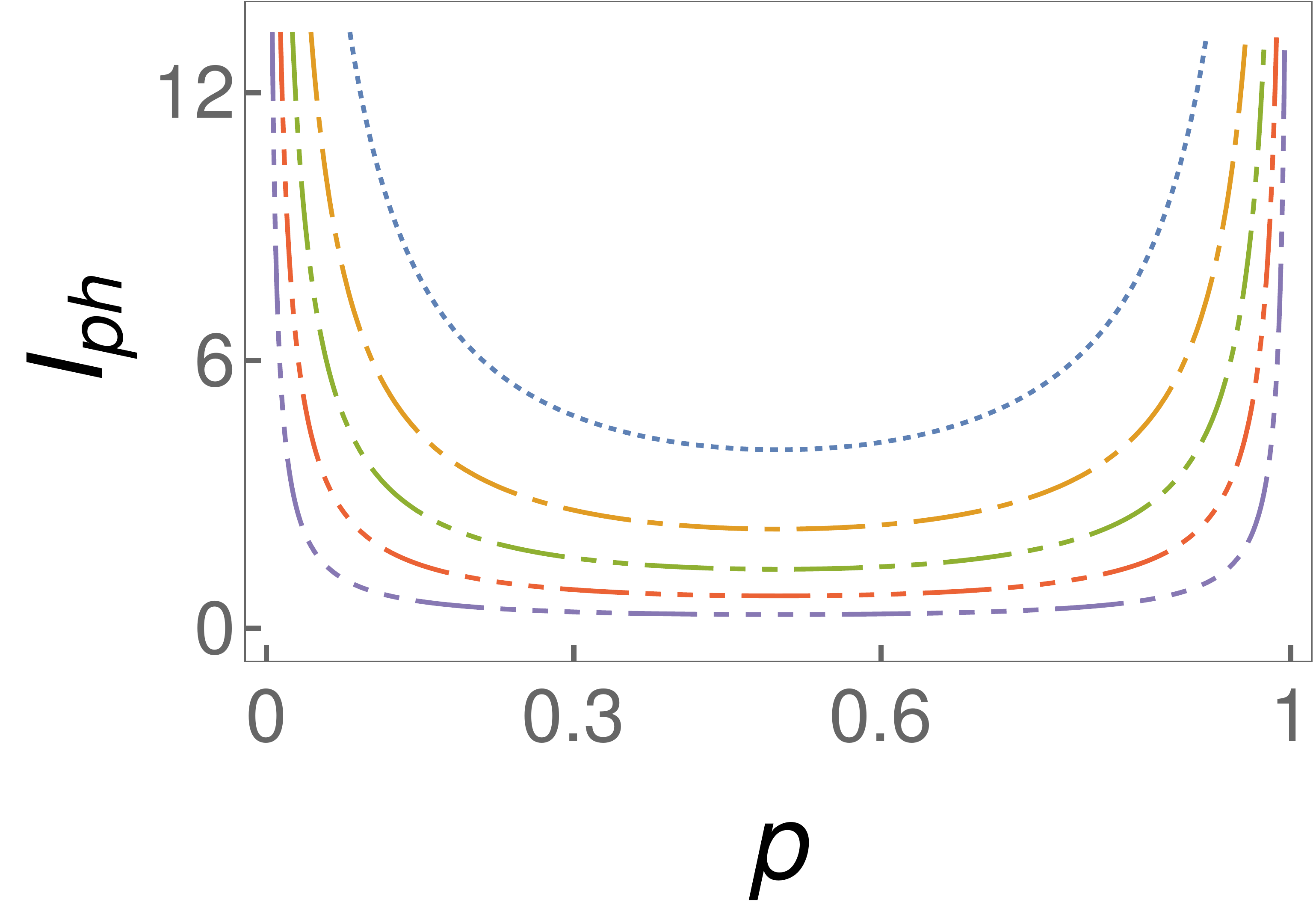}
\caption{\label{fig:QFI_dep_ph_no_loss} QFI
  with no loss of qubits. 
Left figure (depolarizing channel): dotted line: GHZ (optimal strategy); 1-dash line: W-5; 2-dash line: W-10; 3-dash line: W-20; full line: separable state.
Right figure (phase-flip channel): dotted line: GHZ (optimal
separable scheme); 1-dash line: W-5; 2-dash line: W-10; 3-dash line:
W-20; 4-dash line: W-50.} 
\end{figure}

\subsection{ Phase-flip channel }
\subsubsection{GHZ states}
Let 
$\rho_{\mathrm{ph}}^{\text{GHZ-}n}=\mathcal{E}_{\mathrm{ph}}^{(n)}(\rho^{\text{GHZ-}n})$. Applying
the Kraus operators, one obtains immediately
\begin{equation}
\rho_{\mathrm{ph}}^{\text{GHZ-}n}=\frac{1}{2} \left( \dens{0,\mathbb{0}_n}{0,\mathbb{0}_n} +\dens{1,\mathbb{1}_n}{1,\mathbb{1
}_n}\right)+\frac{1-2p}{2} \left(\dens{1,\mathbb{1}_n}{0,\mathbb{0}_n}+\dens{0,\mathbb{0}_n}{1,\mathbb{1}_n}  \right) \,.
\end{equation}

The QFI for $p$, $I_{\mathrm{ph}}^{\text{GHZ-}n}$, is easily found as
the operator has  rank two. The eigenvalues
$\sigma^{\mathrm{ph}}_i$ and eigenvectors $\ket{s^{\mathrm{ph}}_i}$ of
$\rho_{\mathrm{ph}}^{\text{GHZ-}n}$ are 
\begin{equation}
\left\lbrace
\begin{aligned}
&\sigma^{\mathrm{ph}}_1=p\quad, \quad \sigma^{\mathrm{ph}}_2=1-p\\
&\ket{s^{\mathrm{ph}}_1} = \frac{1}{\sqrt{2}}\left( \ket{{0,\mathbb{0}_n}} - \ket{{1,\mathbb{1}_n}} \right) \; , \; \ket{s^{\mathrm{ph}}_2} = \frac{1}{\sqrt{2}}\left( \ket{{0,\mathbb{0}_n}} + \ket{{1,\mathbb{1}_n}} \right) \,.
\end{aligned}\right.
\end{equation}
The eigenvectors are independents of $p$, which means
that the QFI has just the (classical) contribution from the eigenvalues,
\begin{equation}
I_{\mathrm{ph}}^{\text{GHZ-}n}=\frac{1}{p(1-p)} = I_{\mathrm{ph}}^{\mathrm{opt}}\,.
\end{equation}
We see that the QFI for a $(n+1)$-qubit GHZ state used to estimate a
phase-flip channel is independent of the number of ancillas and is
equal to  the optimal QFI achieved by using either a separable state
or a Bell state (but requires more resources in terms of qubits). 

\subsubsection{W states}
The state after acting with the phase-flip channel on the W states,
$\rho_{\mathrm{ph}}^{\text{W-}n}\equiv\mathcal{E}_{\mathrm{ph}}^{(n)}(\rho^{\text{W-}n})$,
is given by
\begin{equation}
\rho_{\mathrm{ph}}^{\text{W-}n}=\frac{1}{n+1} \left( \sum_{i=1}^{n+1} \dens{1_i}{1_i} + \sum_{\substack{ i,j=2 \\ i \neq j }}^{n+1} \dens{1_i}{1_j} \right) +\frac{1-2p}{n+1}\left( \sum_{i=2}^{n+1}( \dens{1_1}{1_i} + \dens{1_i}{1_1} )\right)\,.
\end{equation}
The matrix representation of this state in the computational basis
admits a direct sum decomposition (block structure of the matrix) with
a single non-zero block, of the general form  $K^{(n+1)}(a,b,a)$,
where $a=\frac{1}{n+1}$ and $b=\frac{1-2p}{n+1}$.   

Using eq.(\ref{eq:vp_K}) and the normalized version of
eq.(\ref{eq:vecp_K}) we can compute the QFI, 
\begin{equation}
I_{\mathrm{ph}}^{\text{W-}n}=\frac{4n}{(1+n)^2}\frac{1}{p(1-p)} =\frac{4n}{(1+n)^2}I_{\mathrm{ph}}^{\mathrm{opt}}\,.\label{eq:qfiWph}
\end{equation}
This result shows that the QFI decreases with
 increasing number of ancillas in the W state (see right plot in
 Fig.\ref{fig:QFI_ph_loss_1}): 
 In agreement with the known result on optimality, % we also verify that here 
 the prefactor
 $f(n)=\frac{4n}{(1+n)^2}$ satisfies $f(n)\le 1$ for $n\ge 1$. 
When $n$ goes to
 infinity, $f(n)$ tends to zero, leading to vanishing QFI. 
Even though 
   our analysis is restricted to $n\ge 1$, for $n=0$ the
 W state reduces to $\ket{1}$ which has vanishing QFI such that
 eq.\eqref{eq:qfiWph} is still correct. 

In Fig.\ref{fig:QFI_dep_ph_no_loss} we plot the QFI for
depolarizing and phase-flip channel as a function of $p$ when no
qubits are lost. We 
see that the GHZ states gives the highest QFI (we do not have to
specify the number of ancillas in the GHZ states since it does not
change the QFI). For the W states we observe the decrease of the QFI
when increasing the  number of ancillas, and the convergence either to the
performance of the
separable strategy for the depolarizing channel, or to zero for the
phase-flip channel.

%%%%%%%%%%%%%%%%%%%%%%%%%%%%%%%%%%%%%%%%%%%%
\section{Loosing particles} 
%%%%%%%%%%%%%%%%%%%%%%%%%%%%%%%%%%%%%%%%%%%%
In the first part we looked at the QFI for GHZ and W states in
  the ideal situation of no particle loss in order to check how far we
  are from the optimal case. We now investigate the effect of loosing
  particles.

\subsection{General considerations}
Consider a general extended quantum channel
$\mathcal{E}_{\mathrm{ext}}=\mathcal{E}_P\otimes\idc_A$ 
acting on $\rho$ as
\begin{equation}
\mathcal{E}_{\mathrm{ext}}(\rho)=\sum_k E_k \rho E_k^\dagger =\sum_k (F_k \otimes \id_A) \, \rho \, (F_k^\dagger \otimes \id_A) \,.
\end{equation}
We use subscripts $P$ and $A$ for probe (the first system) and
ancilla, respectively. 

We model the loss of one of the systems by tracing it out
\emph{after} applying the channel. 
Physically it means that the state undergoes properly the
channel, and that after this and before the measurement, one of the
systems is lost. We denote the state which underwent the channel
evolution 
$\mathcal{E}_{\mathrm{ext}}$ and then the loss of the probe as
$ \rho_{A}^{\mathcal{E}_{\mathrm{ext}}}$. Direct calculation shows
that in all generality 
\begin{equation}
\rho_{A}^{\mathcal{E}_{\mathrm{ext}}}\equiv
\trpart{P}{\mathcal{E}_{\mathrm{ext}}(\rho)}=\trpart{P}{\rho}\,,
\end{equation}
the reduced initial state of the ancilla.
In this case there is nothing left to estimate: we cannot get any
information on the extended channel by measuring only the ancilla. 

If it is the ancilla that  is lost after the application of the extended channel
on the initial state we have
\begin{equation}
\rho_{P}^{\mathcal{E}_{\mathrm{ext}}}\equiv\trpart{A}{\mathcal{E}_{\mathrm{ext}}(\rho)}=\mathcal{E}_P(\trpart{A}{\rho})\,.\label{eq:tr_ancilla}
\end{equation}
In this case, loosing the ancilla after extending the channel is
equivalent to starting with the non-extended channel acting on the
reduced state of the probe. Loosing the probe after
applying the 
channel, or starting with an initial state which already suffered the
loss of the probe is hence equivalent. From this point of view, our
subsequent study
amounts to considering new initials states.

\subsection{Loss of one ancilla with a GHZ state}
 When tracing out a qubit from the GHZ state we end up with the mixed
state
\begin{equation}
\trpart{1}{\rho^{\text{GHZ-}n}}=\rho^{\text{GHZ-}n}_1=\left(\ddens{0,\mathbb{0}_{n-1}}+\ddens{1,\mathbb{1}_{n-1}}\right)/2\,.\label{ghzmix}
\end{equation}
Consider first the depolarizing channel.
We are interested in the  QFI $I_{\mathrm{dep},1}^{\text{GHZ-}n}$ of
the state $ 
\rho_{\mathrm{dep},1}^{\text{GHZ-}n}\equiv \trpart{1}{
  \mathcal{E}_{\mathrm{dep}}^{(n)} (\rho^{\text{GHZ-}n}) }$. The
subscript ''1'' on the states, on the trace, and on the QFI indicates that we lost
\emph{one} ancilla. 
In virtue of eq.\eqref{eq:tr_ancilla}) we can also write the state $ \rho_{\mathrm{dep},1}^{\text{GHZ-}n}$ as
\begin{align}\label{eq:rho_dep_1}
 \rho_{\mathrm{dep},1}^{\text{GHZ-}n}&= \mathcal{E}_{\mathrm{dep}}^{(n-1)} ( \trpart{1}{\rho^{\text{GHZ-}n}}) \\
&= \frac{2-p}{4}\left( \ddens{0,\mathbb{0}_{n-1}} +\ddens{1,\mathbb{1}_{n-1}}  \right) \nonumber\\
&\relphantom{=}{} +\frac{p}{4}\left( \ddens{1,\mathbb{0}_{n-1}} +\ddens{0,\mathbb{1}_{n-1}}  \right)\,.\label{eq:rho_dep_1DB}
\end{align}
%Eq.\eqref{eq:rho_dep_1DB} is valid only for $n>1$. 
For $n=1$
the state has only rank two, and is actually the totally mixed state
of one qubit, \emph{which is a stationary state of the depolarizing
  channel and thus leads to a vanishing QFI}.
\begin{equation}
I_{\mathrm{dep},1}^{\text{GHZ-}1}=0\,.
\end{equation}

From eq.\eqref{eq:rho_dep_1DB} we obtain directly the QFI for $n>1$,
\begin{equation}\label{eq:qfi_dep_GHZ_lost1}
I_{\mathrm{dep},1}^{\text{GHZ-}n}=\frac{1}{p(2-p )}\,,
\end{equation}
which is the same QFI as for the non-extended
channel applied to a pure state, $I_{\mathrm{dep}}^{\mathrm{sep}}$. 
This means that instead of starting
with a pure state of a single qubit, we can also start with 
the mixed state \eqref{ghzmix}
and use the extended channel.  

For the phase-flip channel we have
$\rho_{\mathrm{ph},1}^{\text{GHZ-}n}\equiv\trpart{1}{
  \mathcal{E}_{\mathrm{ph}}^{(n)} (\rho^{\text{GHZ-}n})
}=\mathcal{E}_{\mathrm{ph}}^{(n-1)} (\rho^{\text{GHZ-}n) }_1)$. But the
mixed state  
$\rho^{\text{GHZ-}n }_1$ is a stationary state of
$\mathcal{E}_{\mathrm{ph}}^{(n-1)}$, and thus there is nothing to
estimate, $I_{\mathrm{ph},1}^{\text{GHZ-}n}=0$.

\begin{figure}
\includegraphics[scale=0.25]{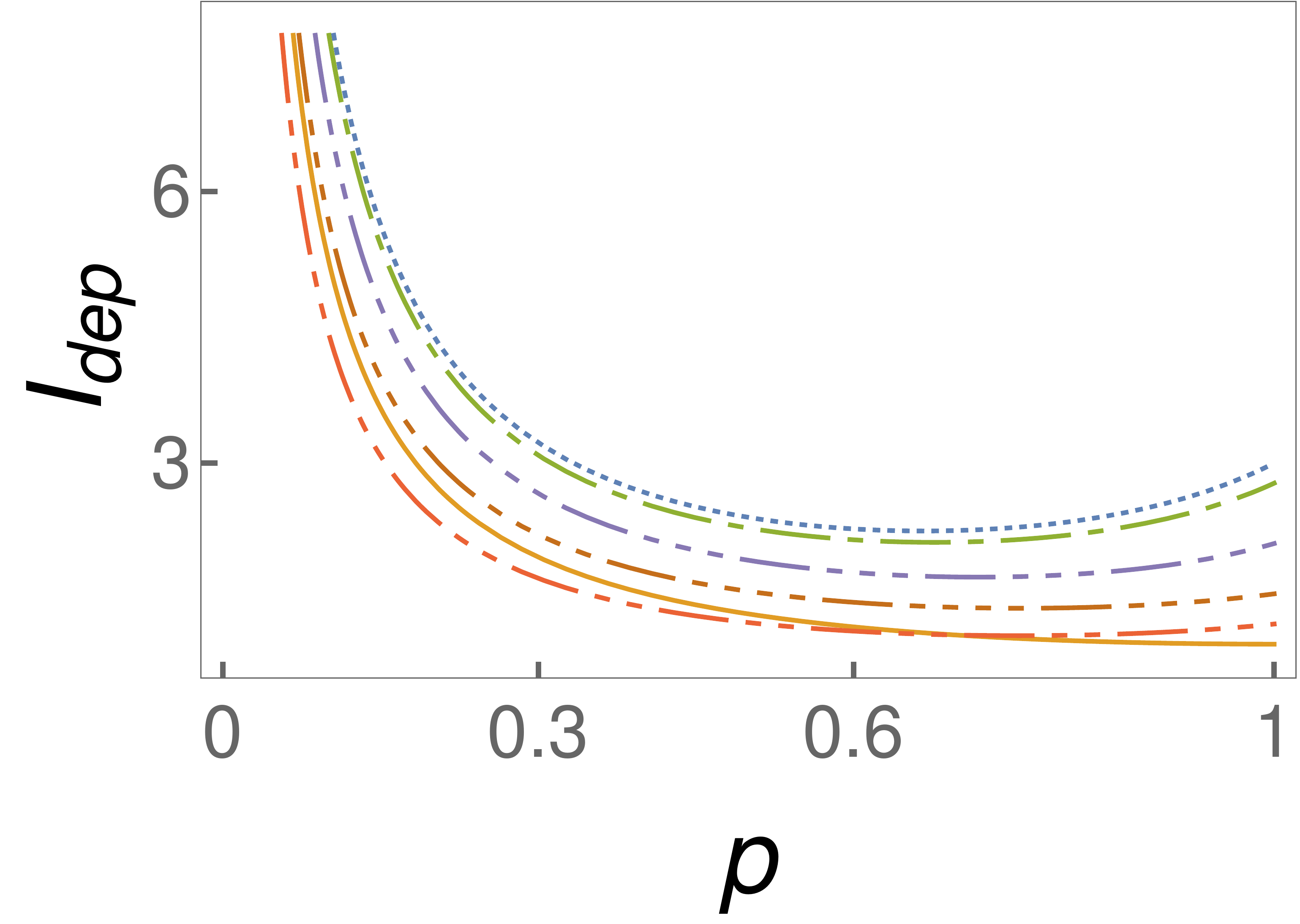}
\includegraphics[scale=0.25]{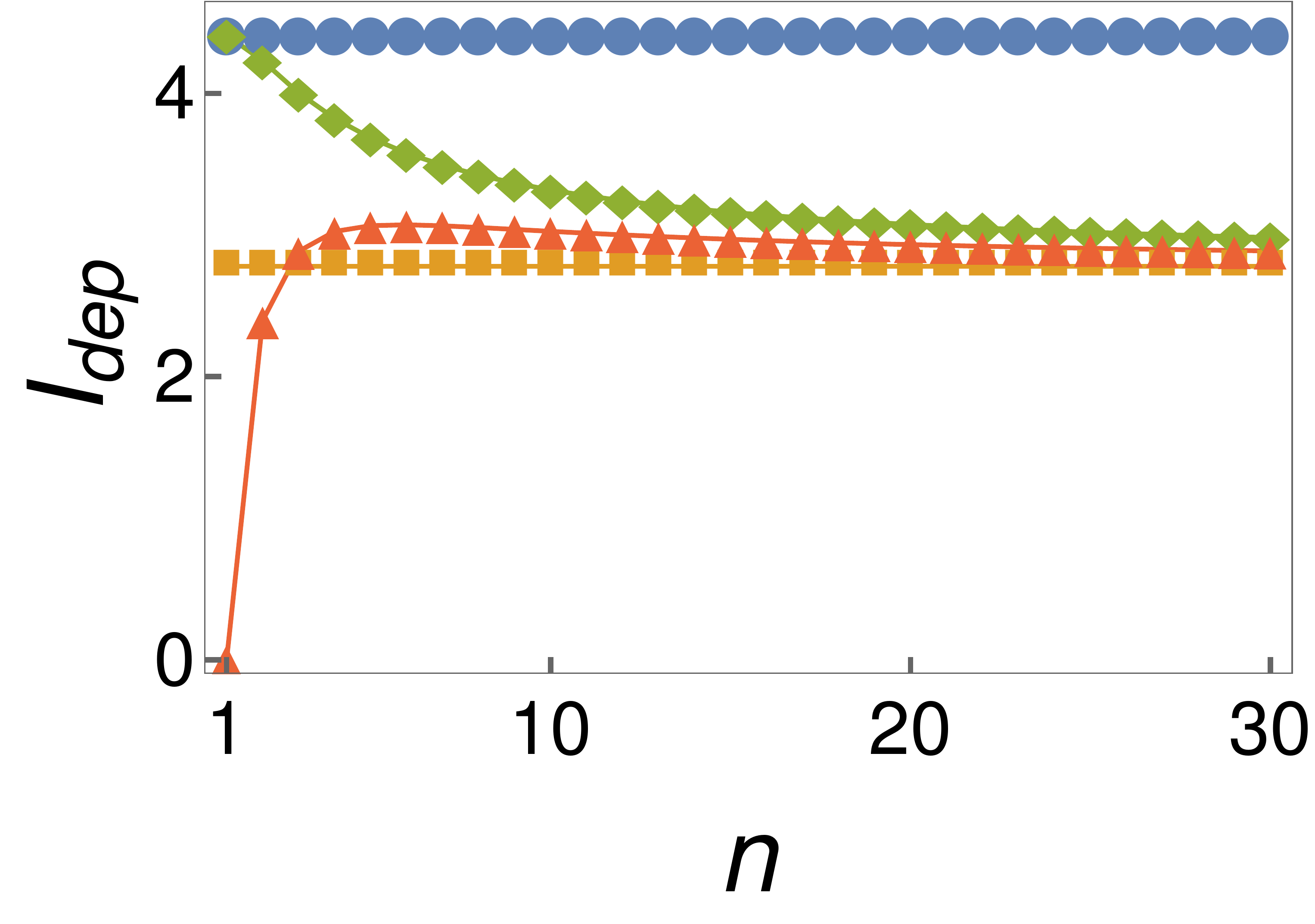}
\caption{Effect of the loss of one ancilla qubit on the QFI for the depolarizing channel. Left figure: dotted 
  line: optimal strategy / GHZ with no loss; 1-dash line: W-2 with no loss; 2-dash
  line: W-2 with one lost; 3-dash line: W-5 with no loss; 4-dash
  line: W-5 with one lost; full line: separable scheme / GHZ with one
  qubit lost. Right figure ($p=0.2$): full circles: GHZ with no loss;
 diamonds: W states with no loss; triangle up: W states with one
 ancilla lost; squares: separable scheme / GHZ with one ancilla
 lost.}
\label{fig:QFI_dep_loss_1}
\end{figure}

\begin{figure}
\includegraphics[scale=0.25]{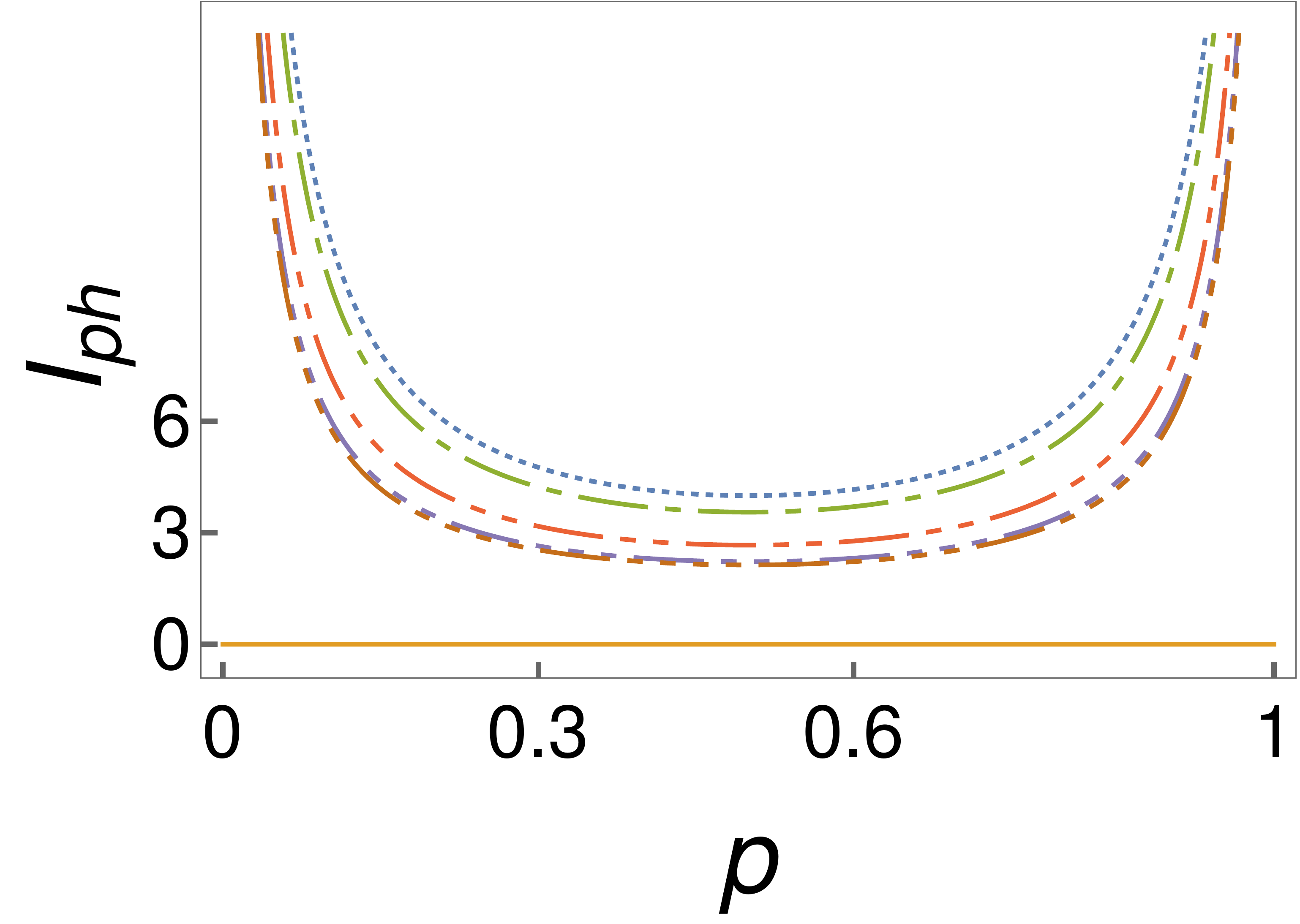}
\includegraphics[scale=0.25]{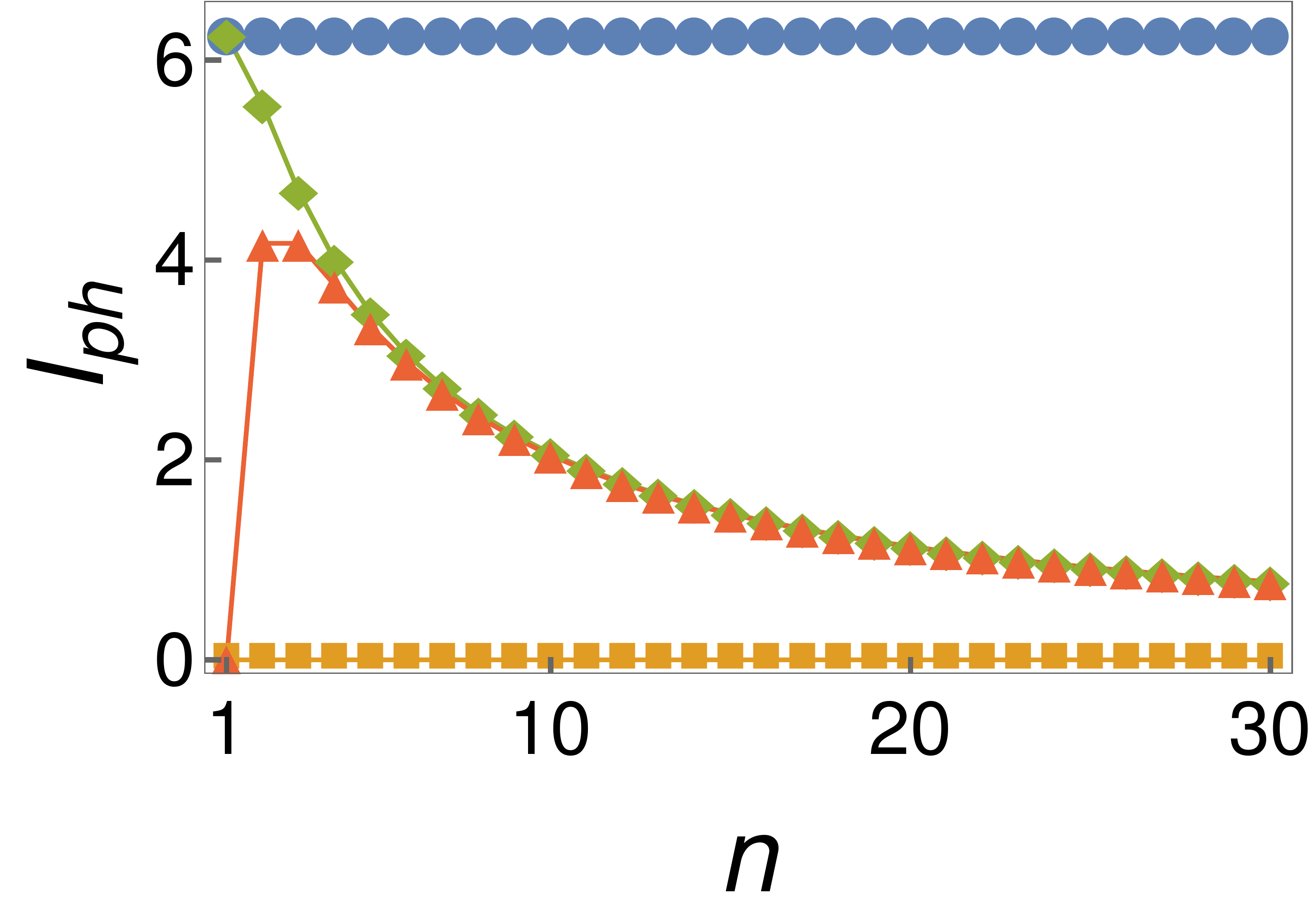}
\caption{Effect of the loss of one ancilla qubit on the QFI for the phase-flip channel.
  Left figure: dotted
  line: GHZ with no loss / optimal separable scheme; 1-dash line: W-2 with no loss; 2-dash
  line: W-2 with one qubit lost; 3-dash line: W-5 with no loss; 4-dash
  line: W-5 with one qubit lost; full line: GHZ with one
  qubit lost.  Right figure ($p=0.2$): full circles: GHZ with no
 loss / 
 optimal separable scheme; diamonds: W states with no loss; triangle
 up: W states with one ancilla lost; squares: GHZ with one ancilla
 lost.} 
\label{fig:QFI_ph_loss_1}
\end{figure}

\subsection{Loss of one ancilla with a W state}
After propagation of a W state through the extended depolarizing channel and the
subsequent loss of an ancilla qubit, the state of the system 
\begin{align}
\rho_{\mathrm{dep},1}^{\text{W-}n}&\equiv\trpart{1}{
  \mathcal{E}_{\mathrm{dep}}^{(n)} (\rho^{\text{W-}n}) } \nonumber \\
&=\frac{1}{n+1}\left( \ddens{0,\mathbb{0}_{n-1}} + \ddens{1_1}\right)+ \frac{2-p}{2(n+1)}\sum_{i,j=2}^n \dens{1_i}{1_j} \nonumber
\\& \relphantom{=}{}+\frac{1-p}{n+1} \sum_{i=2}^n (\dens{1_1}{1_i}+\dens{1_i}{1_1}) + \frac{p}{2(n+1)}\sum_{i,j=2}^n \dens{1,1_i}{1,1_j}\,,
\end{align}
has a block
structure with three non-vanishing blocks: 
\begin{itemize}
\item a first non-contributing $1\times 1$ block composed  by the
  eigenvalue   $\frac{1}{n+1}$. 
\item a second block $G^{(n-1)}(a)$ with $a=\frac{p}{2(n+1)}$.
\item a third block $K^{(n)}(a,b,c)$with $a=\frac{2-p}{2(n+1)}$,
  $b=\frac{1-p}{n+1}$ and $c=\frac{1}{n+1}$. 
\end{itemize} 
 
 This leads to
\begin{equation}\label{eq:qfi_dep_w_n_lost1}
I_{\mathrm{dep},1}^{\text{W-}n}=\frac{n-1}{n+1} \frac{ n (2 p -3)-9}{ p (2 p -3)
   (2 n- p (n-1) )}\,.
\end{equation}

We show in Fig.\ref{fig:QFI_dep_loss_1} the effect of the loss of one
ancilla when estimating the depolarizing channel. In the left figure
the QFI is represented as a function of $p$ for GHZ states with and
without loss, and also for W-2 and W-5 with and without
loss. We see that although W-2 is more efficient than
W-5  in the ideal case, when one qubit is lost W-5 
fairs better than W-2. In the right plot we represent the QFI as a
function of the number of initial  ancillas. We see that by increasing
$n$ the two curves representing the W states with one ancilla lost and the W
states without loss 
converge to the QFI achieved with the separable strategy. 

For the phase-flip channel followed by the loss of one ancilla, the
state of the system is 
$ \rho_{\mathrm{ph},1}^{\text{W-}n}\equiv\trpart{1}{
  \mathcal{E}_{\mathrm{ph}}^{(n)} (\rho^{\text{W-}n}) }=
\mathcal{E}_{\mathrm{ph}}^{(n-1)} ( \trpart{1}{\rho^{\text{W-}n}}) $. 
Direct calculation gives
\begin{equation}
\rho_{\mathrm{ph},1}^{\text{W-}n}=\frac{1}{n+1} \ddens{0,\mathbb{0}_{n-1}}+\frac{n}{n+1} \rho_{\mathrm{ph}}^{\text{W-}(n-1)}\,.
\end{equation}
Note that $\rho_{\mathrm{ph},1}^{\text{W-}n}$   can be written as a
direct sum. Since the first block does
not depend on $p$, we can compute the QFI directly from the second block,
\begin{equation}
I_{\mathrm{ph},1}^{\text{W-}n}=\frac{n}{n+1}I_{\mathrm{ph}}^{\text{W-}(n-1)}=\frac{4(n-1)}{n(n+1)}\frac{1}{p(1-p)}\,.
\end{equation}
We see in the left of Fig.\ref{fig:QFI_ph_loss_1} that this time W-2
has a larger QFI than W-5 
with \emph{and without loss}. In the right figure we observe the
convergence of the QFI to zero for the W states in the ideal case and
with one ancilla lost.

\subsection{Generalization to the loss of  $l$ ancillas}

Now we consider the situation where we loose $l$ ancillas,  $1\le l \leq n$.
Since this loss channel acts only on the ancilla space  it
commutes with the channel acting on the probe, and the situation is
equivalent to starting with a state which underwent already the loss
of the ancillas. 
\subsubsection{Loosing $l$  ancillas in a GHZ state}
When one starts with a GHZ state and looses $l$ qubits, the state becomes
\begin{equation}
\trpart{l}{\rho^{\mathrm{GHZ,n}}}=(\ddens{0,\mathbb{0}_{n-l}}+
\ddens{1,\mathbb{1}_{n-l}})/2\,. %\added a 1/2!
\end{equation}
Loosing one ancilla or $l\ge 2$ ancillas makes no
difference for the QFI. Indeed the GHZ state is so
 sensitive to loss of particles that 
loosing one qubit or more always leads to a mixed state of the same
form (see Sec.\ref{intro}). We thus have for the depolarizing channel and $1\le l \leq n-1$
\begin{equation}
 I^{\text{GHZ-}n}_{\mathrm{dep},l} = I^{\mathrm{sep}}_{\mathrm{dep}}  =\frac{1}{p(2-p)} \label{eq:qfi_dep_GHZ_lostl}\,, 
\end{equation}
and for $n=l$
\begin{equation}
 I^{\text{GHZ-}l}_{\mathrm{dep},l} = 0\,.
\end{equation}
For the phase-flip channel  and $1\le l \leq n$,
\begin{equation}
 I^{\text{GHZ-}n}_{\mathrm{ph},l}=0  \label{eq:qfi_ph_GHZ_lostl}\,.
\end{equation}

\subsubsection{Loosing $l$  ancillas in a W state}
For the W state, the situation is substantially different since the
form of the 
state depends on the number of lost ancillas: 
\begin{equation}
\rho^{\text{W-}n}_l=\trpart{l}{\rho^{\text{W-}n}}=\frac{l}{n+1}
\ddens{0,\mathbb{0}_{n-l}} +\frac{n+1-l}{n+1} \rho^{\text{W-}(n-l)}\,. \label{rla}
\end{equation}

For the depolarizing channel, the state 
\begin{align}
\rho_{\mathrm{dep},l}^{\text{W-}n}&\equiv \trpart{l}{\rho^{\text{W-}n}_{\mathrm{dep}}} \nonumber \\
&=\frac{2l-p(l-1)}{2(n+1)} \ddens{0,\mathbb{0}_{n-1}} + \frac{2+p(l-1)}{2(n+1)} \ddens{1_1}+ \frac{2-p}{2(n+1)}\sum_{i,j=2}^{n+1-l} \dens{1_i}{1_j} \nonumber
\\&\relphantom{=}{}+\frac{1-p}{n+1} \sum_{i=2}^{n+1-l} (\dens{1_1}{1_i}+\dens{1_i}{1_1}) + \frac{p}{2(n+1)}\sum_{i,j=2}^{n+1-l} \dens{1,1_i}{1,1_j}\,,
\end{align}
 has three non-vanishing
blocks: 
\begin{itemize}
\item a first  $1\times 1$ block composed  by the eigenvalue $\frac{2l-p(l-1)}{2(n+1)}$.
\item a second block $G^{(n-l)}(a)$ with $a=\frac{p}{2(n+1)}$.
\item a third block $K^{(n+1-l)}(a,b,c)$with $a=\frac{2-p}{2(n+1)}$, $b=\frac{1-p}{n+1}$ and $c=\frac{2+p(l-1)}{2(n+1)}$.
\end{itemize} 
They lead to the QFI
\begin{multline}\label{eq:qfi_dep_W_lostl}
 I^{\text{W-}n}_{\mathrm{dep},l}=\Big\lbrace -2 p  \Big(-2 (l (l+2)-1) n^2+l (l
   (3 l+2)-9) n+l (l (3 l+4)-9)+8 n+2\Big)+ \\  (l-1)
   (l+3) (n+1) p^2 (2 l-n-1)+4 l (l+2) (n+3)
   (l-n)\Big\rbrace \\  \times \frac{1}{ (n+1) p ((l-1) p -2 l) ((l+3) p
  -2 (l+2)) (l (2-2 p )+(n+1) (p -2))}\,.
\end{multline}

\begin{figure}
\centering \includegraphics[scale=0.5]{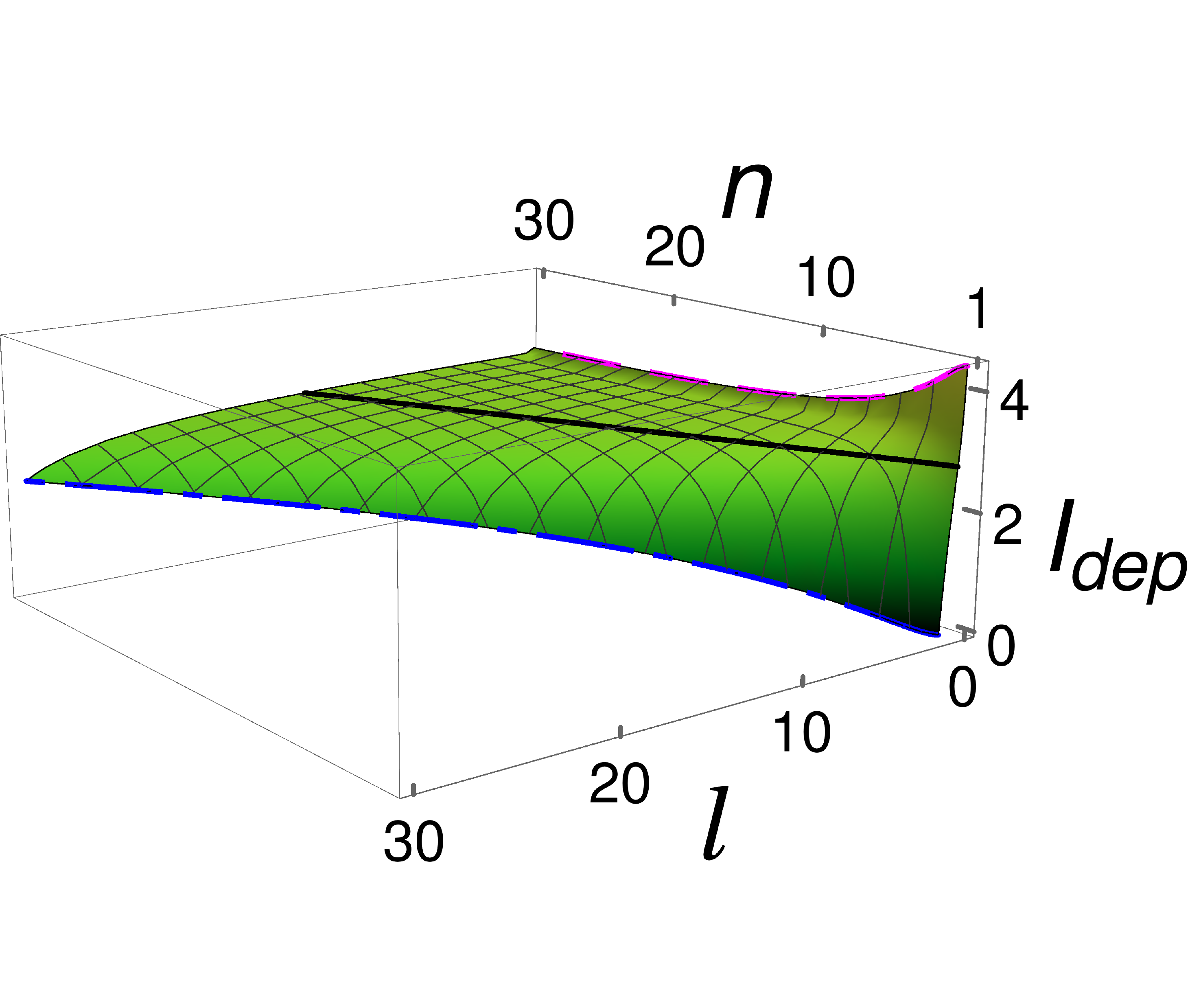}
\caption{QFI for the  depolarizing channel for a W state with $n$
  ancillas and $l$ lost ones ($p=0.2$). The full line corresponds to the separable strategy or GHZ with loss, the dashed line to the case where no ancillas are lost ($l=0$) and the 1-dash line to the case where all ancillas are lost ($l=n$).}\label{fig:DepWnl}
\end{figure}

One checks that by setting $l$ to zero or to one we recover our
previous results \eqref{eq:qfi_dep_w_n} and
\eqref{eq:qfi_dep_w_n_lost1}, respectively.  
In terms of gain due to extension, we can calculate the number of lost
ancillas as a function of the number of initial ancillas such that
the scheme stays more efficient than the separable strategy. This
function is cumbersome but actually behaves mainly linearly with a
slope of $0.5$. This means that when more than half the
ancillas are lost, the strategy of using W states becomes less efficient than the
separable strategy. But for 
depolarizing channel,  this strategy is equivalent to the use of a GHZ
state with some ancillas lost
(\ref{eq:qfi_dep_GHZ_lost1},\ref{eq:qfi_dep_GHZ_lostl}). Thus this
bound gives us also the value of $l$ as a function of $n$ for which it
is worth to start with a GHZ state rather than with a W state. This 
is visualized in  Fig.\ref{fig:DepWnl}, representing the QFI
for the depolarizing channel as a function of $n$ and $l$, and where
the full black line represents the  QFI for the separable strategy or GHZ with loss.

\begin{figure}
\includegraphics[scale=0.24]{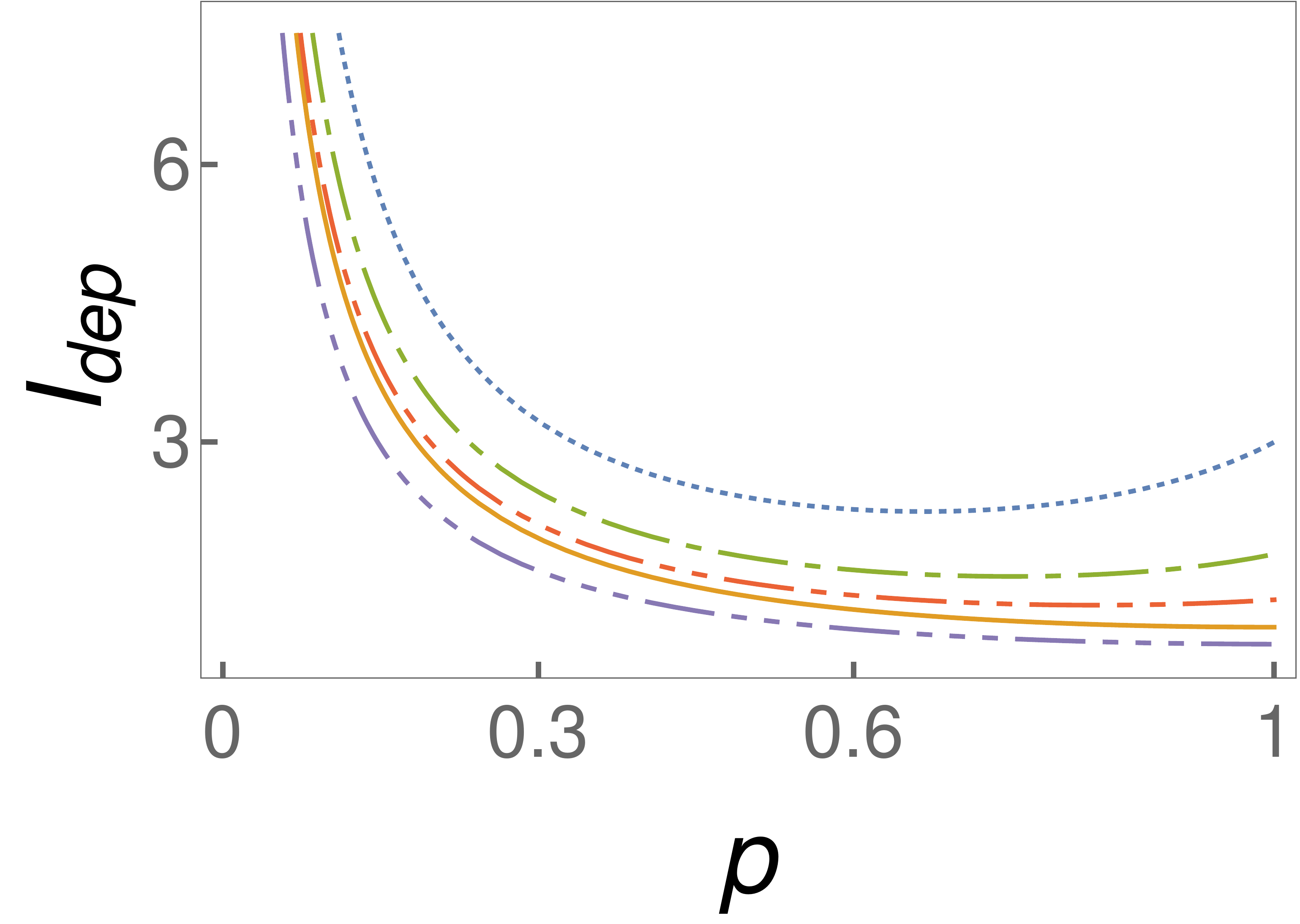}
\includegraphics[scale=0.26]{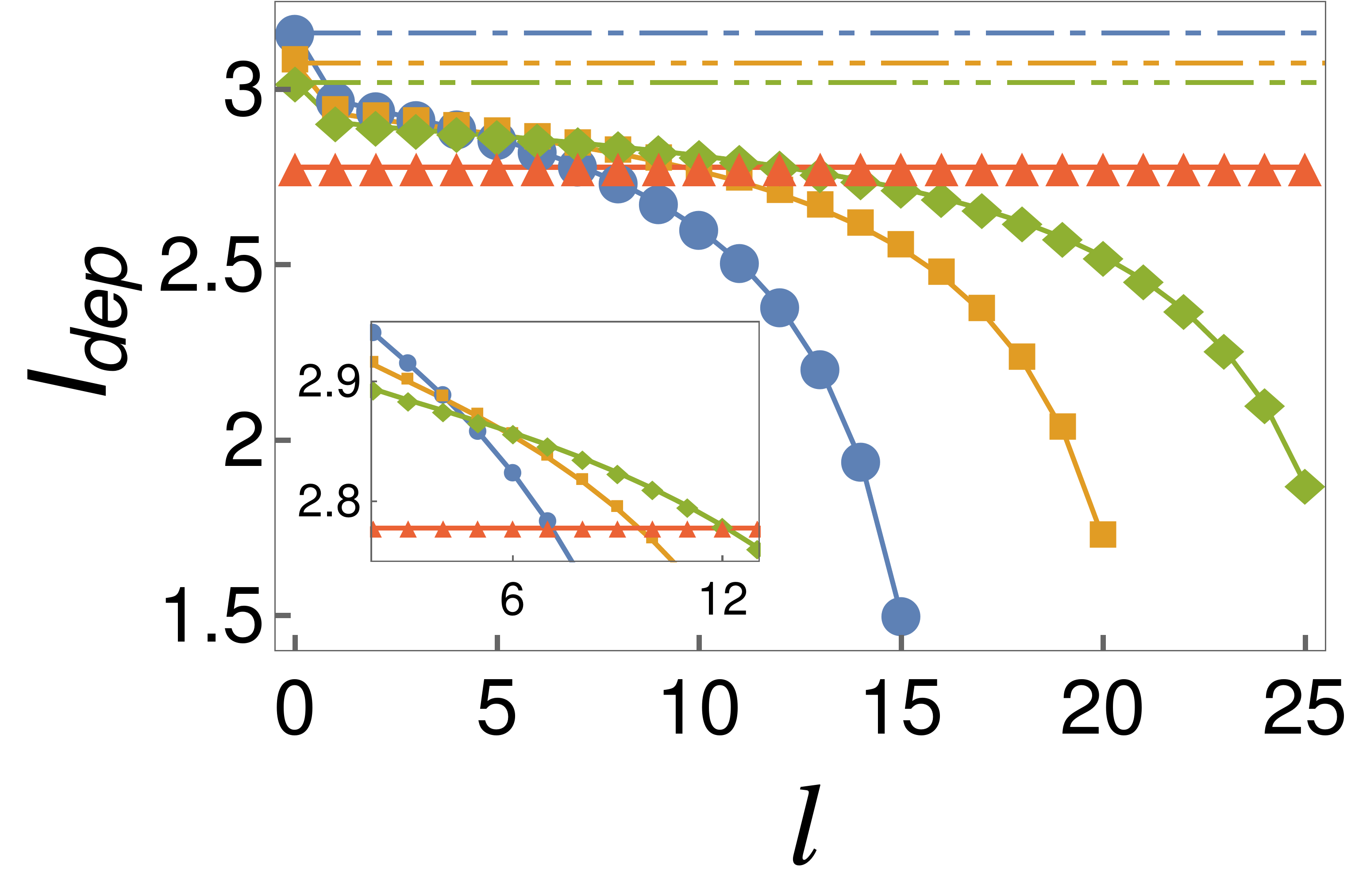}
\caption{QFI for depolarizing channel for
  arbitrary loss. Left figure: dotted line: GHZ with no loss; 1-dash
  line: W-8 with no loss; 2-dash line: W-8 with 2 lost; 3-dash
  line: W-8 with 6 lost; full line: separable strategy  / GHZ with
  one lost. 
Right figure ($p=0.2$):  1-dash line: W-15 with no loss; full circles: W-15
with loss; 2-dash line: W-20 with no loss;  squares: W-20 with loss;
3-dash line: W-25 with no loss;  diamonds: W-25 with loss: triangle
up: separable strategy  / GHZ with one lost. 
}\label{fig:QFI_dep_l_lost} 
\end{figure}

In Fig.\ref{fig:QFI_dep_l_lost} we demonstrate the effect of the loss on the
estimation of the depolarizing channel. In the left plot we show the
QFI as a function of $p$. We plot the optimal result (dotted line) and
the separable strategy (full line). The different dash lines show
W-8 with either no loss, or  two, or six ancillas
lost. In agreement with the bound discussed in the previous paragraph, for six ancillas lost in W-8 , the protocol is less efficient than the
separable one / GHZ with loss. In the right plot we show the QFI for the
depolarizing channel as a function of the number of lost ancillas. We
observe that W 
states with a larger number of ancillas are more resistant to the
loss of qubits, but have a lower initial QFI. There is
a compromise for the optimal choice of $n$ in a W state between
initial QFI and robustness to the loss. When the number of lost ancillas equals roughly half the number of
initial ancillas, the W states become less efficient than the GHZ
states (this is more clear in the subplot). When all the ancillas are
lost, the QFI still not vanishes, provided that $n>1$: Setting $l$ to $n$ in
eq.\eqref{eq:qfi_dep_W_lostl}, leads to 
\begin{equation}
 I^{\text{W-}n}_{\mathrm{dep},n}=\frac{(n-1)^2}{(2+(n-1)p)(p+n(2-p))}\,,
\end{equation}
which converges to $I^{\mathrm{sep}}_{\mathrm{dep}}$ when $n$ goes to
infinity. For $n=1$,  $I^{\text{W-}1}_{\mathrm{dep},1}=0$.

In the case of the phase-flip channel, the state \eqref{rla}
after application of the channel and
 loss of $l$ ancillas has the form of a direct sum involving a known state, leading to the
QFI
 \begin{equation}
 I^{\text{W-}n}_{\mathrm{ph},l}=\frac{n+1-l}{n+1}  I^{\text{W-}(n-l)}_{\mathrm{ph}}=\frac{4(n-l)}{(n+1)(n+1-l)}\frac{1}{p(1-p)}\,.
 \end{equation}
As expected, the QFI decreases  as function of $l$:  The more ancillas
are lost the worse is the estimation. When all ancillas are lost
the QFI vanishes, since the resulting state is insensitive to the
phase-flip channel. 

\begin{figure}
\includegraphics[scale=0.24]{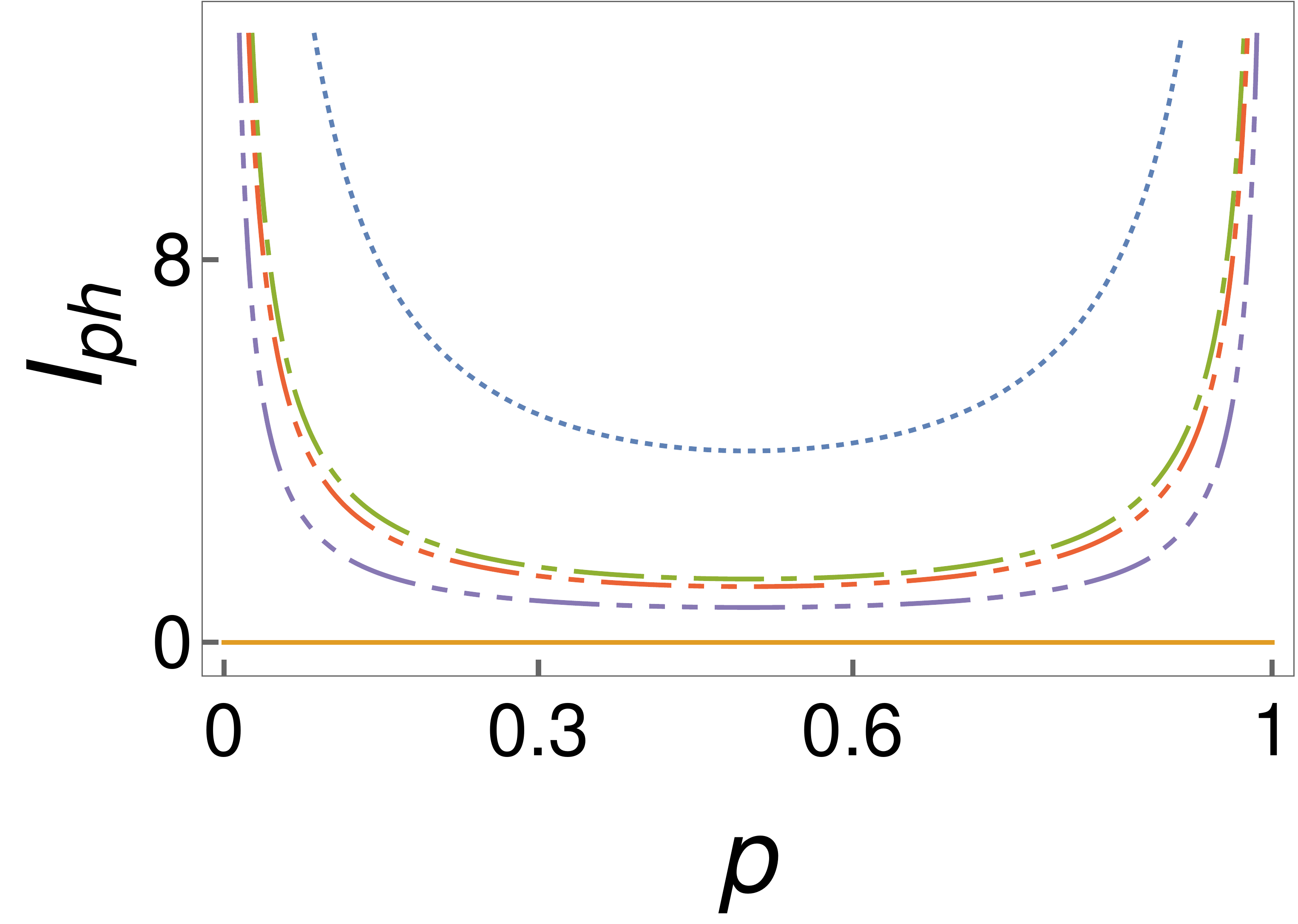}
\includegraphics[scale=0.26]{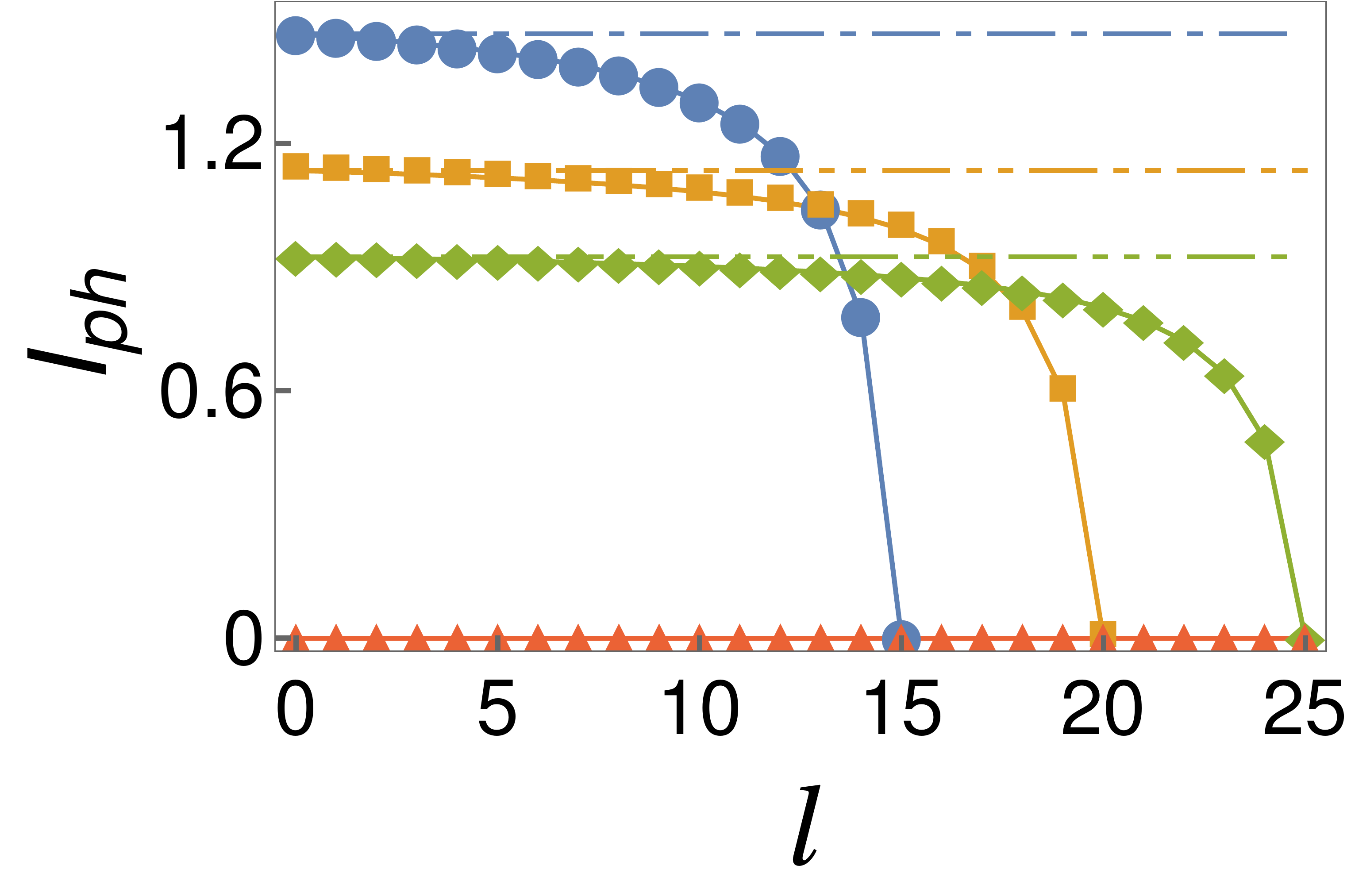}
\caption{QFI for phase-flip channel for the loss of an arbitrary
  number of qubits. 
 Left figure: dotted line: optimal strategy / GHZ with no loss; 1-dash line: W-10 with no loss; 2-dash line: W-10 with 6 lost; 3-dash line: W-10 with 9 lost; full line: GHZ with at least one ancilla lost.
Right figure  ($p=0.2$): 1-dash line: W-15 with no loss; full circles:
 W-15
with loss; 2-dash line: W-20 with no loss;  squares: W-20 with loss;
3-dash line: W-25 with no loss;  diamonds: W-25 with loss: triangle 
up: GHZ with loss (GHZ without loss is not
represented). }\label{fig:QFI_ph_l_lost} 
\end{figure}

This is demonstrated in Fig.\ref{fig:QFI_ph_l_lost}. The left
plot shows the QFI as a function of $p$. In the
                               right plot $p=0.2$,
 and we plot the QFI as a
function of the number of lost ancillas for W states. 
The more ancillas we add the smaller the initial QFI, but also the QFI
decreases more 
slowly as function of $l$. This leads to an optimal number of initial
ancillas for a given number of ancillas lost, even though we have
 to remember that for the phase-flip channel the best strategy is to not
 use any ancillas at all (see Sec.\ref{sec.benchmark}).

\subsection{Gain \emph{versus} robustness}
There is a competition between the initial
value of the QFI and the robustness for W states  for both channels
(although for the phase-flip channel we know that the optimal scheme
is the non-extended one). 
 
\begin{figure}
\includegraphics[scale=0.26]{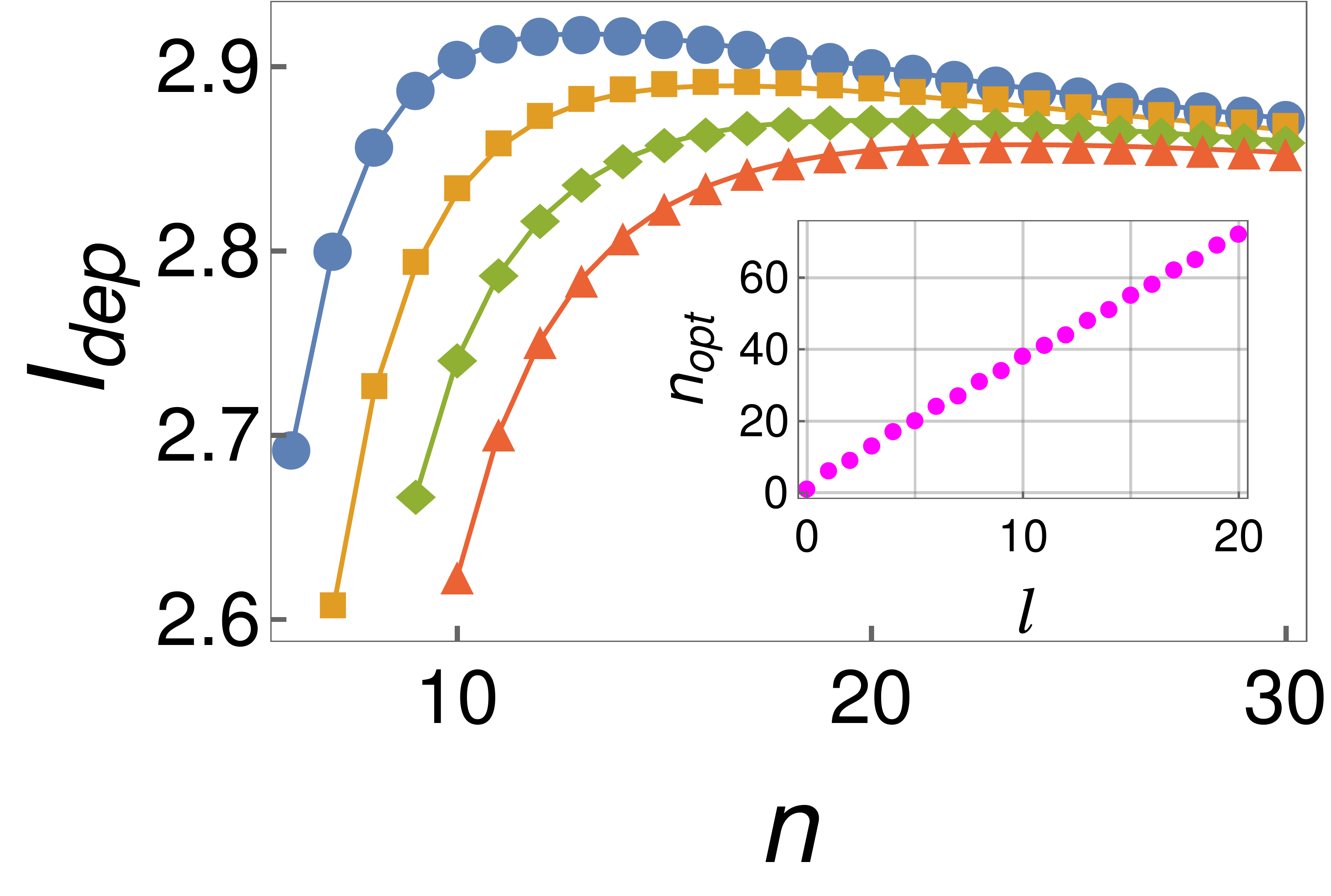}
\includegraphics[scale=0.26]{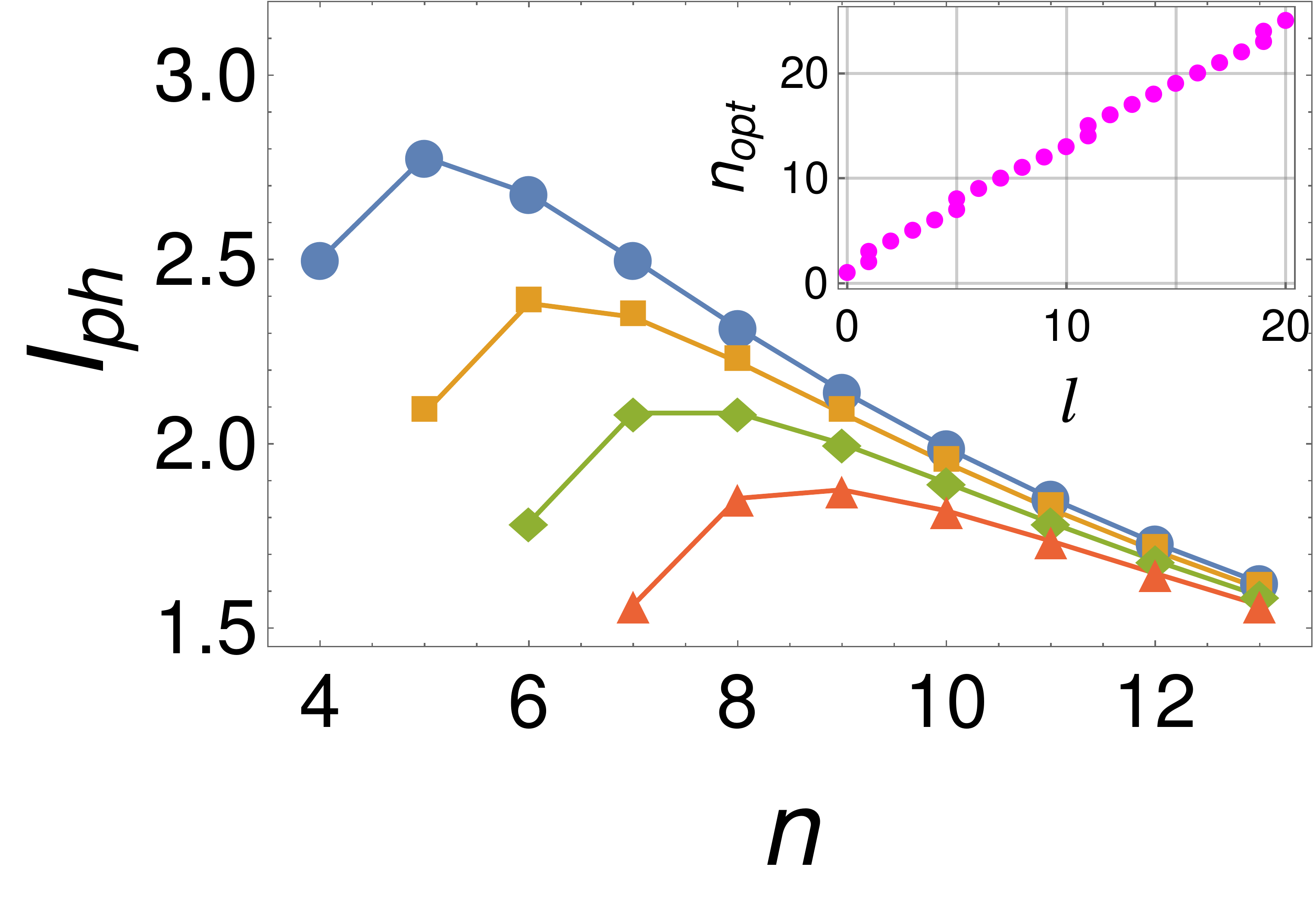}
\caption{Main plots: QFI as a function of the
  number of initial ancillas in a W state for a fixed number of lost
  ancillas. The full circles 
correspond to three ancillas lost, the
  squares to four, the diamonds to five, and the triangle to six. In
  the insets we see the optimal number of initial ancillas in a W
  state as a function of the number of lost ancillas. The left plot
  corresponds to the depolarizing channel, the right one
  to the phase-flip channel ($p=0.2$). }\label{fig:n_opt} 
\end{figure}

For the depolarizing channel, when looking at the right plot in Fig. \ref{fig:QFI_dep_l_lost}, we see that while in the ideal case ($l=0$) W-15 is more efficient than W-25, this is already no longer true when six ancillas are lost as the inset clearly shows. More generally there exists for a given fixed number $l$ of  ancillas
lost an optimal number $n_\mathrm{opt,dep}(l)$ of initial ancillas in the W
state, see left plot of 
Fig.\ref{fig:n_opt}. The function $n_\mathrm{opt,dep}(l)$ has a
complicated form, but its leading term  is given by 
\begin{equation}
n_{\mathrm{opt,dep}}(l)\simeq\left( 2+\frac{2}{\sqrt{2-p}} \right)l \,,
\end{equation}
 which for $p=0.2$ gives roughly $3.5 l$. We see that this is in good
 agreement with the inset of the left plot of
 Fig.\ref{fig:n_opt}. Nevertheless, when increasing the number of
 ancillas in the W state we get a QFI closer to the one of the
 separable strategy, and thus the small gain in QFI may not justify
 the use of so many ancillas. As an example, when loosing fifteen
 ancillas, the best W state is  W-55 (the leading term in this case
 will give $n_{\mathrm{opt,dep}}=52$ or  53), but its QFI equals 2.81
 and  the QFI for the separable strategy equals 2.77. 
 
A similar behavior is observed for the phase flip
channel. Although there the optimal strategy consists to not add any
ancilla, the study of the QFI for a fixed number of lost ancillas
leads also to a maximum as represented in the right plot in
Fig.\ref{fig:n_opt}. We can here too calculate the optimal number of
initial ancillas as a function of lost ancillas in a W state 
\begin{equation}
n_{\mathrm{opt,ph}}(l)=
\left\lbrace\begin{aligned} 
& \lfloor l+\sqrt{1+l} \rfloor \equiv l_\mathrm{f} \quad \text{ if } \quad  I^{\text{W-}l_\mathrm{f}}_{\mathrm{ph},{l}} > I^{\text{W-}l_\mathrm{c}}_{\mathrm{ph},{l}} \,, \\ 
&\lceil l+\sqrt{1+l}  \rceil  \equiv l_\mathrm{c}\quad \text{ if } \quad  I^{\text{W-}l_\mathrm{f}}_{\mathrm{ph},{l}} < I^{\text{W-}l_\mathrm{c}}_{\mathrm{ph},{l}} \,, \\
&\lbrace l_\mathrm{f},l_\mathrm{c}  \rbrace\quad \text{ if } \quad  I^{\text{W-}l_\mathrm{f}}_{\mathrm{ph},{l}} = I^{\text{W-}l_\mathrm{c}}_{\mathrm{ph},{l}}\,,
 \end{aligned}\right. 
\end{equation}
with $\lfloor \rfloor$ the floor function and $\lceil \rceil$ the
ceiling function. Thus $n_\mathrm{opt,ph}$  scales roughly linearly with $l$.

\section{Conclusions}
We investigated the robustness of channel estimation schemes for
depolarizing and phase-flip channels of qubits extended to ancilla
qubits, when one or several of the qubits can get lost.  Without loss
of qubits, the optimal estimation strategy consists for both channels in
extending the channel by a single ancilla qubit that remains
untouched, but feeding the whole channel with a maximally entangled
state
\cite{fujiwara_quantum_2003}. 

For the depolarizing channel this leads, when no qubit is lost, to a
real improvement compared to the non-extended case. For the phase-flip channel
the maximum quantum Fisher information (QFI) can also be achieved with
a non-extended channel fed with a specific pure state, showing that no extension is necessary.  
We extended this investigation to the case
where an arbitrary number of qubits can be added or lost, including
the original probe qubit. We used GHZ and W states as input states for
the channels.  

For the GHZ states, the QFI in the absence of loss is equal to the
optimal one for both channels, independently of the number of
ancillas. In the presence of loss, for the depolarizing channel and
provided that not all the ancillas are lost --- in which case the QFI
vanishes ---, the QFI is independent of the number of lost ancillas
and equals the one of the non-extended case. For the phase flip
channel the loss of already one ancilla leads to a vanishing QFI. 

For the W states, the QFI for the depolarizing channel without loss
decreases with the number of added ancillas. While for one ancilla we
are in the optimal case, when the number of ancillas goes to infinity
the QFI goes to the QFI of the separable strategy. The interesting
point lies in the fact that the W states are more resistant to loss,
as for a fixed number of lost ancillas, there always exists a W state
with 
larger QFI after the loss of these ancillas than the one of the
separable strategy. In this sense the 
W states, although  not optimal without loss of qubits, 
can lead to a better estimation in 
non-ideal situations for the depolarizing channel. The resistance to
loss is also observed in the 
phase-flip channel, but does not lead to any improvement in
estimation, since it is still better to not add ancillas at all.  

{\bf Acknowledgments:} 
We gratefully acknowledge useful correspondence with
  A.~Fujiwara.

\section{Appendix: decomposition of matrices}
In order to calculate the QFI we need to diagonalize the density
matrix. For the states in which we are interested, there are two
matrices $K^{(m)}(a,b,c)$ and  $G^{(m)}(a)$ that recurrently appear in
the block decomposition of the states:
\begin{itemize}
\item The $m \times m$ matrix $K^{(m)}(a,b,c)$:
\begin{equation}
K^{(m)}(a,b,c)=
\begin{pmatrix}
a &\cdots& a & b \\ 
\vdots  & \ddots &\vdots & \vdots \\ 
a&  \cdots & a& b \\ 
b &  \cdots & b &c
\end{pmatrix} \,.
\end{equation}
 This matrix has rank two, the two eigenvalues 
 \begin{equation}\label{eq:vp_K}
 \lambda^{(K)}_\pm=\frac{1}{2}\left( c+a(m-1)\pm \sqrt{\left(c-a(m-1)\right)^2+4b^2(m-1)}\right) \,,
 \end{equation}
and the two corresponding non-normalized eigenvectors
 \begin{equation}\label{eq:vecp_K}
 \mathbf{v}^{(K)}_\pm=(2b,\cdots,2b, Y^{(K)}_\pm)\,,
 \end{equation}
 with $Y^{(K)}_\pm=c-a(m-1)\pm \sqrt{\left(c-a(m-1)\right)^2+4b^2(m-1)}$.

\item The $m \times m$ matrix $G^{(m)}(a)$:
\begin{equation}
G^{(m)}(a)=\begin{pmatrix}
a & \cdots& a  \\ 
\vdots&\ddots&\vdots\\ 
a& \cdots & a \\ 
\end{pmatrix} \,,
\end{equation}
which only non-zero eigenvalue is \begin{equation}\label{eq:vp_G}
 \lambda^{(G,m)}=m a \,,
\end{equation} 
and the  non-normalized corresponding eigenvector is
 \begin{equation}\label{eq:vecp_G}
 \mathbf{v}^{(G,m)}(a)=(1,\cdots,1)\,.
\end{equation}
\end{itemize}

%\bibliographystyle{plain}
%\bibliography{../channel_identification,../../mybibs_bt}
%\bibliography{../../mybibs_bt}
\bibliography{mybibs_bt}
%\nocite{*}

\end{document}